\documentclass[useAMS,usenatbib]{mn2e}

\usepackage{graphicx}

\newcommand{\pc}{\, {\rm pc} }

\newcommand{\km}{\, {\rm km} }
\newcommand{\Msun}{\ensuremath{{\rm M}_\odot}}
\newcommand{\RCR}{R_{\mathrm{CR}}}
\newcommand{\Kpc}{\ \mathrm{kpc}}
\newcommand{\ab}{a_{\mathrm{b}}}
\def\X{\exp [i \, m ( \Omega_P t -  \theta)]}

\title[Radial and vertical flows induced by galactic spiral arms]{Radial and vertical flows induced by galactic spiral arms: likely contributors to our ``wobbly Galaxy''}

\author[C. Faure, A. Siebert \& B. Famaey]
{Carole~Faure$^1$\thanks{carole.faure@astro.unistra.fr},
Arnaud~Siebert$^1$,
Benoit~Famaey$^1$ \\
$^1$Observatoire Astronomique, Universit\'e de Strasbourg, CNRS UMR 7550, France}

\begin{document}

\date{\today}
\maketitle

\begin{abstract}
  In  an equilibrium  axisymmetric  galactic disc,  the mean  galactocentric
  radial and  vertical velocities  are expected to  be zero  everywhere.  In
  recent years, various large  spectroscopic surveys have however shown that
  stars of  the Milky Way disc  exhibit non-zero mean  velocities outside of
  the Galactic plane in both the Galactocentric radial and vertical velocity
  components. While  radial velocity structures  are commonly assumed  to be
  associated  with  non-axisymmetric components  of  the  potential such  as
  spiral  arms or bars,  non-zero vertical  velocity structures  are usually
  attributed to excitations by external  sources such as a passing satellite
  galaxy or  a small  dark matter substructure  crossing the  Galactic disc.
  Here, we use a three-dimensional test-particle simulation to show that the
  global stellar response to a spiral perturbation induces both a radial 
  velocity flow and non-zero vertical motions. The resulting structure of the 
  mean velocity field
  is qualitatively  similar to what is  observed across the  Milky Way disc.
  We show  that such a pattern also naturally emerges from an
  analytic toy model based on  linearized Euler equations.  We conclude that
  an external perturbation of the disc might not be a requirement to explain
  all  of the observed  structures in  the vertical velocity  of stars
  across the Galactic disc. Non-axisymmetric internal perturbations can also
  be the source of the observed mean velocity patterns.
\end{abstract}

\maketitle

\section{Introduction}

The Milky Way has long been  known to possess spiral structure, but studying
the nature  and the  dynamical effects  of this structure  has proven  to be
elusive  for decades.   Even though  its fundamental  nature is  still under
debate today, it has nevertheless started to be recently considered as a key
player in galactic dynamics and evolution (e.g., Antoja et al.~2009; Quillen
et al.~2011; L\'epine  et al.~2011; Minchev et al.~2012;  Roskar et al.~2012
for  recent works,  or Sellwood~2013  for a  review).  However,  zeroth order
dynamical models  of the Galaxy  still mostly rely  on the assumptions  of a
smooth  time-independent  and  axisymmetric  gravitational  potential.   For
instance,  recent  determinations of  the  circular  velocity  at the  Sun's
position  and of  the peculiar  motion of  the Sun  itself all  rely  on the
assumption of  axisymmetry and on minimizing  the non-axisymmetric residuals
in the  velocity field (Reid et  al.~2009; McMillan \& Binney  2010; Bovy et
al.~2012; Sch\"onrich~2012).  Such zeroth  order assumptions are handy since
they  allow  us   to  develop  dynamical  models  based   on  a  phase-space
distribution function depending only on three isolating integrals of motion,
such  as  the  action  integrals  (e.g.,  Binney~2013;  Bovy  \&  Rix~2013).
Actually, an action-based approach does  not necessarily have to rely on the
axisymmetric assumption,  as it  is also possible  to take into  account the
main non-axisymmetric component (e.g., the bar, see Kaasalainen \& Binney 1994) by modelling  the system in
its rotating  frame (e.g., Kaasalainen 1995). However the  other non-axisymmetric components  such as
spiral arms rotating with a different pattern speed should then nevertheless
be treated through perturbations (e.g., Kaasalainen~1994; McMillan~2013).

The main  problem with such  current determinations of  Galactic parameters,
through  zeroth order  axisymmetric models,  is that  it is  not  clear that
assuming axisymmetry and dynamical equilibrium to fit a benchmark model does
not  bias  the  results,  by  e.g.   forcing this  benchmark  model  to  fit
non-axisymmetric features  in the observations  that are not present  in the
axisymmetric model  itself. This  means that the  residuals from  the fitted
model  are   not  necessarily  representative  of  the   true  amplitude  of
non-axisymmetric motions.  In  this respect, it is thus  extremely useful to
explore the full range of possible effects of non-axisymmetric features such
as spiral arms in both fully controlled test-particle simulations as well as
self-consistent simulations, and to compare these with observations.

With  the advent  of  spectroscopic and  astrometric surveys,  observational
phase-space information for stars in an increasingly large volume around the
Sun  have allowed  us to  see more  and more  of these  dynamical  effect of
non-axisymmetric components  emerge in the  data.  Until recently,  the most
striking  features were  found in  the solar  neighbourhood in  the  form of
moving groups, i.e.  local velocity-space  substructures shown to be made of
stars of  very different  ages and chemical  compositions (e.g.,  Chereul et
al. 1998,  1999; Dehnen~1998; Famaey  et al.~2005, 2007, 2008;  Pomp\'eia et
al.~2011).  Various non-axisymmetric  models have been argued to  be able to
represent  these  velocity  structures  equally  well,  using  transient 
(e.g., De Simone et al. 2004) or
quasi-static  spirals (e.g., Quillen \& Minchev 2005; Antoja et al. 2011),  
with or  without  the  help  of the  outer  Lindblad
resonance from  the central bar (e.g., Dehnen 2000; Antoja et al. 2009; Minchev et al. 2010; 
McMillan 2013; Monari et al. 2013). 
The effects  of non-axisymmetric components
have also  been analyzed  a bit  less locally by  Taylor expanding  to first
order the planar velocity field in the cartesian frame of the Local Standard
of Rest, i.e. measuring the Oort constants $A$, $B$, $C$ and $K$ (Kuijken \&
Tremaine~1994; Olling \& Dehnen~2003), a  procedure valid up to distances of
less than 2~kpc. While old data were compatible with the axisymmetric values
$C=K=0$  (Kuijken \& Tremaine~1994),  a more  recent analysis  of ACT/Tycho2
proper  motions  of   red  giants  yielded  $C  =   -10  \,  {\rm  km}\,{\rm
  s}^{-1}\,{\rm  kpc}^{-1}$ (Olling  \&  Dehnen~2003).  Using  line-of-sight
velocities  of 213713 stars  from the  RAVE survey  (Steinmetz et  al. 2006;
Zwitter et  al. 2008; Siebert  et al. 2011a;  Kordopatis et al.  2013), with
distances $d<2 \,$kpc in the longitude interval $-140^\circ < l < 10^\circ$,
Siebert et al. (2011b) confirmed this  value of $C$, and estimated a value of
$K= +6\,{\rm  km}\,{\rm s}^{-1}\,{\rm kpc}^{-1}$,  implying a Galactocentric
radial velocity\footnote{In  this paper,  {\it 'radial velocity'}  refers to
  the  Galactocentric   radial  velocity,  not  to  be   confused  with  the
  line-of-sight  (l.o.s.)  velocity.}   gradient  of  $C+K  =  \partial  V_R
/  \partial R  \simeq -  4\,{\rm km}\,{\rm  s}^{-1}\,{\rm kpc}^{-1}$  in the
solar   suburb  (extended   solar  neighbourhood,   see  also   Williams  et
al. 2013). The projection onto  the plane of the mean line-of-sight velocity
as a  function of distance  towards the Galactic centre  ($|l|<5^\circ$) was
also examined by Siebert et al. (2011b) both for the full RAVE sample and for
red clump  candidates (with an  independent method of  distance estimation),
and clearly  confirmed that the RAVE  data are not compatible  with a purely
axisymmetric  rotating disc.   This result  is  not owing  to
systematic distance errors  as considered in Binney et  al.  (2013), because
the {\it geometry}  of the radial velocity  flow cannot be reproduced by systematic distance errors alone (Siebert et  al.~2011b; Binney et
al.~2013). Assuming, to  first order, that the observed  radial velocity map
in the solar  suburb is representative of what would  happen in a razor-thin
disc, and that the spiral arms are long-lived, Siebert et al. (2012) applied
the  classical density wave  description of  spiral arms  (Lin \&  Shu 1964;
Binney \&  Tremaine 2008)  to constrain their  parameters in the  Milky Way.
They found that the best-fit  was obtained for a two-armed perturbation with
an amplitude  corresponding to $\sim 15$\%  of the background  density and a
pattern speed $\Omega_P  \simeq 19 \,{\rm Gyr}^{-1}$, with  the Sun close to
the 4:1 inner  ultra-harmonic resonance (IUHR). This result  is in agreement
with studies based on the location  of moving groups in local velocity space
(Quillen \& Minchev~2005; Antoja  et al.~2011; Pomp\'eia et al.~2011).  This
study was advocated to be a useful first order benchmark model to then study
the effect of spirals in three dimensions.

In  three   dimensions,  observations  of  the  solar   suburb  from  recent
spectroscopic surveys  actually look even  more complicated. Using  the same
red  clump giants  from RAVE,  it  was shown  that the  mean {\it  vertical}
velocity  was also  non-zero  and  showed clear  structure  suggestive of  a
wave-like  behaviour (Williams et  al.~2013). Measurements  of line-of-sight
velocities for 11000  stars with SEGUE also revealed  that the mean vertical
motion  of stars  reaches up  to 10~km/s  at heights  of 1.5~kpc  (Widrow et
al.~2012), echoing previous similar results  by Smith et al. (2012). This is
accompanied by a significant wave-like North-South asymmetry in SDSS (Widrow
et al.~2012; Yanny  \& Gardner~2013). Observations from LAMOST  in the outer
Galactic disc  (within 2~kpc  outside the Solar  radius and 2~kpc  above and
below the Galactic  plane) also recently revealed (Carlin  et al.~2013) that
stars  above the  plane  exhibit a  net  outward motion  with downward  mean
vertical  velocities,  whilst stars  below  the  plane  exhibit the  opposite
behaviour in terms of vertical  velocities (moving upwards, i.e. towards the
plane too), but  not so much in terms of  radial velocities, although slight
differences are  also noted. There is  thus a growing body  of evidence that
Milky Way disc  stars exhibit velocity structures across  the Galactic plane
in {\it  both} the  Galactocentric radial and  vertical components.  While a
global radial velocity gradient such as  that found in Siebert et al. (2011b)
can naturally be explained with non-axisymmetric components of the potential
such as spiral arms, such an explanation is {\it a priori} less self-evident
for vertical  velocity structures. For instance, it was recently shown that the central bar 
cannot produce such vertical features in the solar suburb (Monari et al. 2014). 
For  this reason, such  non-zero vertical
motions  are generally  attributed to  vertical excitations  of the  disc by
external means such as a  passing satellite galaxy (Widrow et al.~2012). The
Sagittarius dwarf has been pinpointed as a likely culprit for creating these
vertical density  waves as  it plunged through  the Galactic disc  (Gomez et
al.~2013),  while other  authors  have argued  that  these could  be due  to
interaction of the  disc with small starless dark  matter subhalos (Feldmann
\& Spolyar~2013).

Here, we rather investigate whether such vertical velocity structures can be
expected as the  response to disc non-axisymmetries, especially spiral arms, 
in the absence of external perturbations.  As a  first step in  this direction, we  propose to
qualitatively investigate the response of a typical old thin disc stellar 
population to a spiral perturbation in
controlled test particle  orbit integrations. Such test-particle simulations
have revealed  useful in 2D  to understand the effects  of non-axisymmetries
and their  resonances on the  disc stellar velocity field,  including moving
groups  (e.g., Antoja  et  al.~2009, 2011;  Pomp\'eia  et al.~2011), Oort 
constants (e.g., Minchev et al. 2007), radial
migrations (e.g.,  Minchev \&  Famaey~2010), or the  dip of  stellar density
around   corotation  (e.g.,  Barros   et  al.~2013).   Recent  test-particle
simulations in 3D have rather concentrated on the effects of the central bar
(Monari  et al.~2013, 2014), while  we concentrate  here on  the effect  of spiral
arms, with special attention to mean vertical motions. In Sect. 2, we give
details on  the model potential,  the initial conditions and  the simulation
technique,  while  results  are  presented  in Sect.  3,  and  discussed  in
comparison  with solutions  of linearized  Euler equations.  Conclusions are
drawn in Sect.~4.

\section{Model}

To pursue our  goal, we use a standard test-particle  method where orbits of
massless particles  are integrated in  a time-varying potential.   We start
with an  axisymmetric background potential  representative of the  Milky Way
(Sect.~2.1), and  we adiabatically grow  a spiral perturbation on  it within
$\sim 3.5$~Gyr.  Once  settled, the spiral perturbation is  kept at its full
amplitude.   This  is  not  supposed  to be  representative  of  the  actual
complexity  of  spiral structure  in  real  galaxies, where  self-consistent
simulations indicate  that it  is often  coupled to a  central bar  and/or a
transient  nature with  a lifetime  of the  order of  only a  few rotations.
Nevertheless,  it allows  us to  investigate the  stable response  to  an old
enough  spiral  perturbation   ($\sim  600  \,$Myr  to  $1   \,$Gyr  in  the
self-consistent simulations of Minchev  et al.  2012).  The adiabatic growth
of  this spiral  structure is  not meant  to be  realistic, as  we  are only
interested in  the orbital  structure of the  old thin disk  test population
once the perturbation is stable.

We  generate initial  conditions  for  our test  stellar  population from  a
discrete realization  of a  realistic phase-space distribution  function for
the  thin disc  defined in  integral-space (Sect~2.2),  and  integrate these
initial   conditions  forward   in   time  within   a  given   time-evolving
background+spiral potential  (Sect.~2.3). We then analyze  the mean velocity
patterns  seen in  configuration space,  both radially  and  vertically, and
check  whether such patterns  are stable  within the  rotating frame  of the
spiral.

\subsection{Axisymmetric background potential}

The axisymmetric  part of the Galactic  potential is taken to  be Model~I of
Binney \& Tremaine (2008). Its main parameters are summarized in Table~1 for
convenience. The central bulge has a truncated power-law density of the form
\begin{equation}
\rho_b(R,z) = \rho_{b0} \times \left( \frac{\sqrt{R^2 + (z/q_b)^2}}{a_b} \right)^{-\alpha_b} {\rm exp}\left( -\frac{R^2+(z/q_b)^2}{r_b^2} \right)
\end{equation}
where $R$ is  the Galactocentric radius within the  midplane, $z$ the height
above  the plane,  $\rho_{b0}$  the central  density,  $r_b$ the  truncation
radius, and $q_b$ the flattening. The total mass of the bulge is $M_b = 5.18
\times 10^9 \Msun$.

The stellar  disc is  a sum of  two exponential  profiles (for the  thin and
thick discs):
\begin{equation}
\rho_d(R,z) = \Sigma_{d0} \times \left( \sum_{i=1}^{i=2} \frac{\alpha_{d,i}}{2z_{d,i}} {\rm exp}\left(-\frac{|z|}{z_{d,i}}\right) \right)  {\rm exp}\left(-\frac{R}{R_d}\right)
\end{equation}
where  $\Sigma_{d0}$  is the  central  surface  density, $\alpha_{d,1}$  and
$\alpha_{d,2}$  the relative  contributions  of the  thin  and thick  discs,
$z_{d,1}$  and  $z_{d,2}$  their  respective scale-heights,  and  $R_d$  the
scale-length.  The total  mass of  the disc  is $M_d  = 5.13  \times 10^{10}
\Msun$.The disc potential also includes a contribution from the interstellar
medium of the form
\begin{equation}
\rho_g(R,z) = \frac{\Sigma_g}{2z_g} \times {\rm exp}\left(-\frac{R}{R_g} -\frac{R_m}{R} - \frac{|z|}{z_g}\right)
\end{equation}
where $R_m$ is  the radius within which  there is a hole close  to the bulge
region, $R_g$ is the scale-length, $z_g$ the scale-height, and $\Sigma_g$ is
such  that  it contributes  to  25\%  of the  disc  surface  density at  the
galactocentric radius of the Sun.

Finally, the dark halo is  represented by an oblate two-power-law model with
flattening $q_h$, of the form
\begin{eqnarray}
\rho_{h}(R,z)&=&\rho_{h0} \times \left( \frac{\sqrt{R^2 + (z/q_h)^2}}{a_h} \right)^{-\alpha_h}\times\nonumber\\&&\left( 1 + \frac{\sqrt{R^2 + (z/q_h)^2}}{a_h} \right)^{\alpha_h - \beta_h}.
\end{eqnarray}

\begin{table}
  \caption{Parameters of the axisymmetric background model potential (Binney \& Tremaine~2008)}
  \label{tab:potaxi}
  \centering
  \begin{center}
    \begin{tabular}{ll} \hline \hline Parameter & Axisymmetric potential \\ \hline
      $M_b(\Msun)$ & $5.18 \times 10^9$ \\
      $M_d(\Msun)$ & $5.13 \times 10^{10}$ \\
      $M_{h, <100 \Kpc}(\Msun)$ & $6. \times 10^{11}$ \\
      $\rho_{b0}(\Msun \pc^{-3})$ & $0.427$\\
      $\ab(\Kpc)$ & $1.$ \\
      $r_b(\Kpc)$ & $1.9$ \\
      $\alpha_b$ & $1.8$ \\     
      $q_b$ & $0.6$ \\
      $\Sigma_{d0}+\Sigma_g(\Msun \pc^{-2})$ & 1905.\\
      $R_d(\Kpc)$ & 2. \\
      $R_g(\Kpc)$ & 4. \\
      $R_m(\Kpc)$ & 4. \\
      $\alpha_{d,1}$ & $14/15$ \\
      $\alpha_{d,2}$  & $1/15$ \\
      $z_{d,1}(\Kpc)$ & 0.3 \\
      $z_{d,2}(\Kpc)$ & 1. \\
      $z_g(\Kpc)$ & 0.08 \\
      $\rho_{h0}(\Msun \pc^{-3})$ & 0.711 \\
      $a_h(\Kpc)$ & $3.83$ \\
      $\alpha_h$ & $-2.$ \\
      $\beta_h$ & $2.96$ \\
      $q_h$ & $0.8$ \\\hline
    \end{tabular}
  \end{center}
\end{table}

\begin{figure}
\includegraphics[width=7cm]{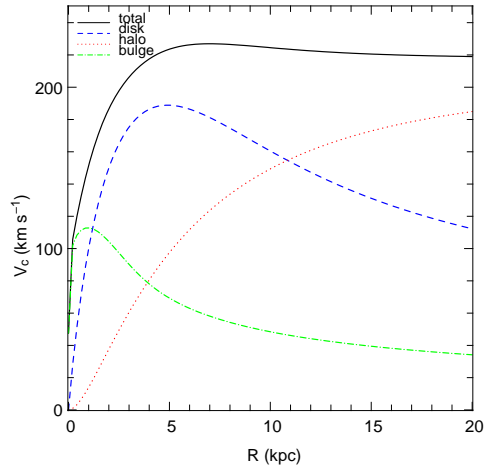}
\caption{Rotation curve corresponding to the background axisymmetric potential}
\label{f:rc}
\end{figure}

The  potential   is  calculcated  using   the  GalPot  routine   (Dehnen  \&
Binney~1998).    The  rotation  curve   corresponding  to   this  background
axisymmetric potential  is displayed in Fig.~\ref{f:rc}.   For radii smaller
than $11$~kpc, the total rotation curve (black line) is mostely infuenced by
the disc (blue dashed line) and above by the halo (red dotted line).

\subsection{Initial conditions}

The  initial conditions  for  the test  stellar  population are  set from  a
discrete  realization  of  a  phase-space distribution  function  (Shu~1969,
Bienaym\'e \& S\'echaud~1997) which can be written in integral space as:

\begin{equation}
f(E_R,L_z,E_z)=\frac{\Omega \, \rho_d}{\sqrt{2} \kappa \pi^{\frac{3}{2}} \sigma^2_R \sigma_z} \exp \left( \frac{-(E_R-E_c)}{\sigma^2_R}-\frac{E_z}{\sigma^2_z} \right)
\end{equation}
in  which the  angular  velocity $\Omega$,  the  radial epicyclic  frequency
$\kappa$ and  the disc density  in the plane  $\rho_d$ are all  functions of
$L_z$, being taken at the radius $R_c(L_z)$ of a circular orbit of angular
momentum $L_z$. The scale-length of the disc is taken to be 2~kpc as for the
background potential.  The energy $E_c(L_z)$  is the energy of  the circular
orbit of angular momentum $L_z$ at the radius $R_c$. Finally, the radial and
vertical  dispersions $\sigma^2_R$  and  $\sigma^2_z$ are  also function  of
$L_z$ and are expressed as:
\begin{equation}
\sigma^2_R=\sigma^2_{R_\odot}\exp\left( \frac{2R_{\odot}-2R_c}{R_{\sigma_R}}\right),
\end{equation}
\begin{equation}
\sigma^2_z=\sigma^2_{z_\odot}\exp\left( \frac{2R_{\odot}-2R_c}{R_{\sigma_z}}\right)
\end{equation}
where  $R_{\sigma_R}/R_d  =  R_{\sigma_z}/R_d  = 5$.  The  initial  velocity
dispersions thus decline exponentially with radius but at each radius, it is
isothermal as a  function of height. These initial values are  set in such a
way as to be representative of the  old thin disc of the Milky Way after the
response to the  spiral perturbation. Indeed, the old thin  disc is the test
population we want to investigate the response of.

\begin{figure}
\includegraphics[width=4cm]{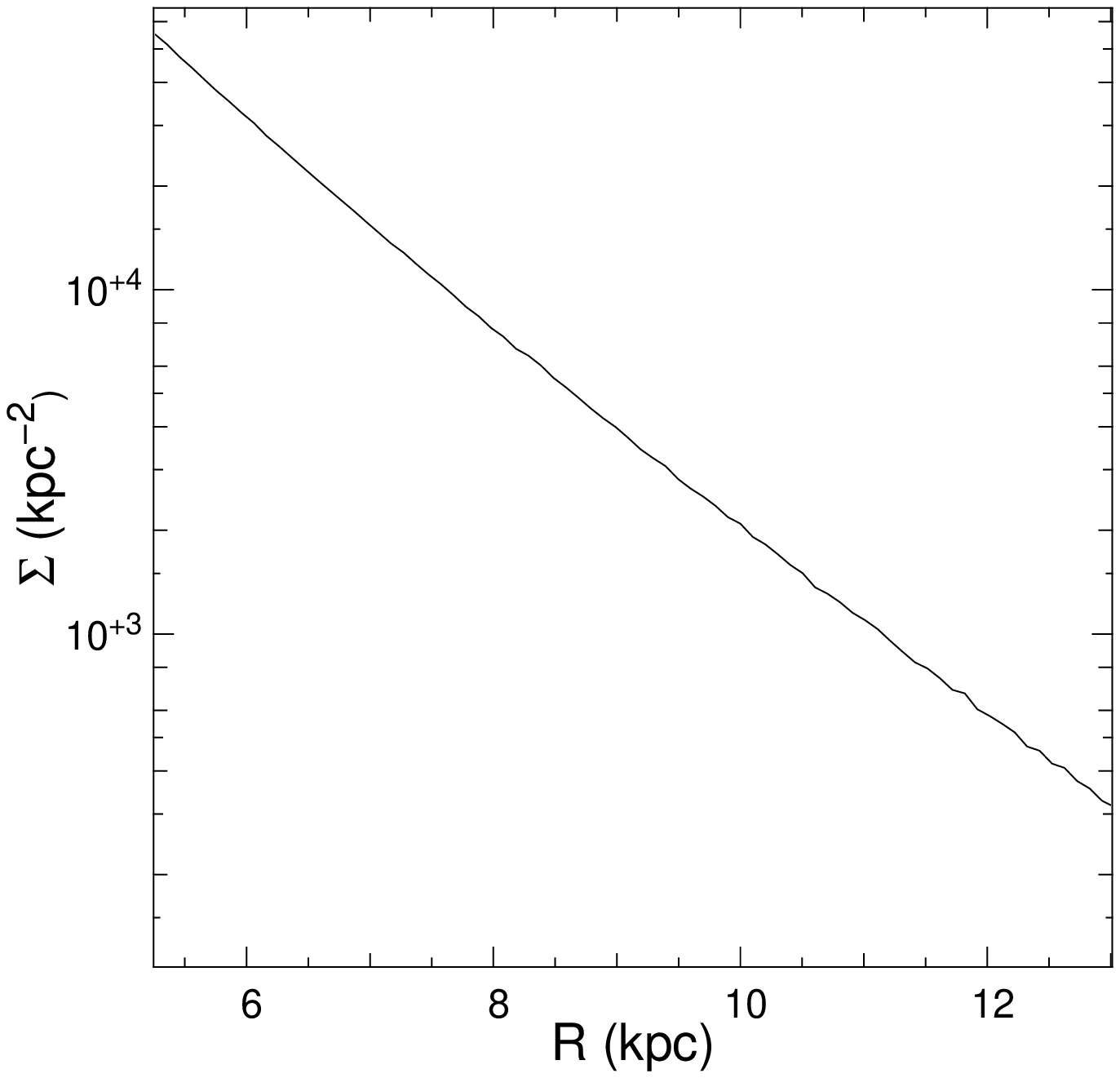}
\includegraphics[width=4cm]{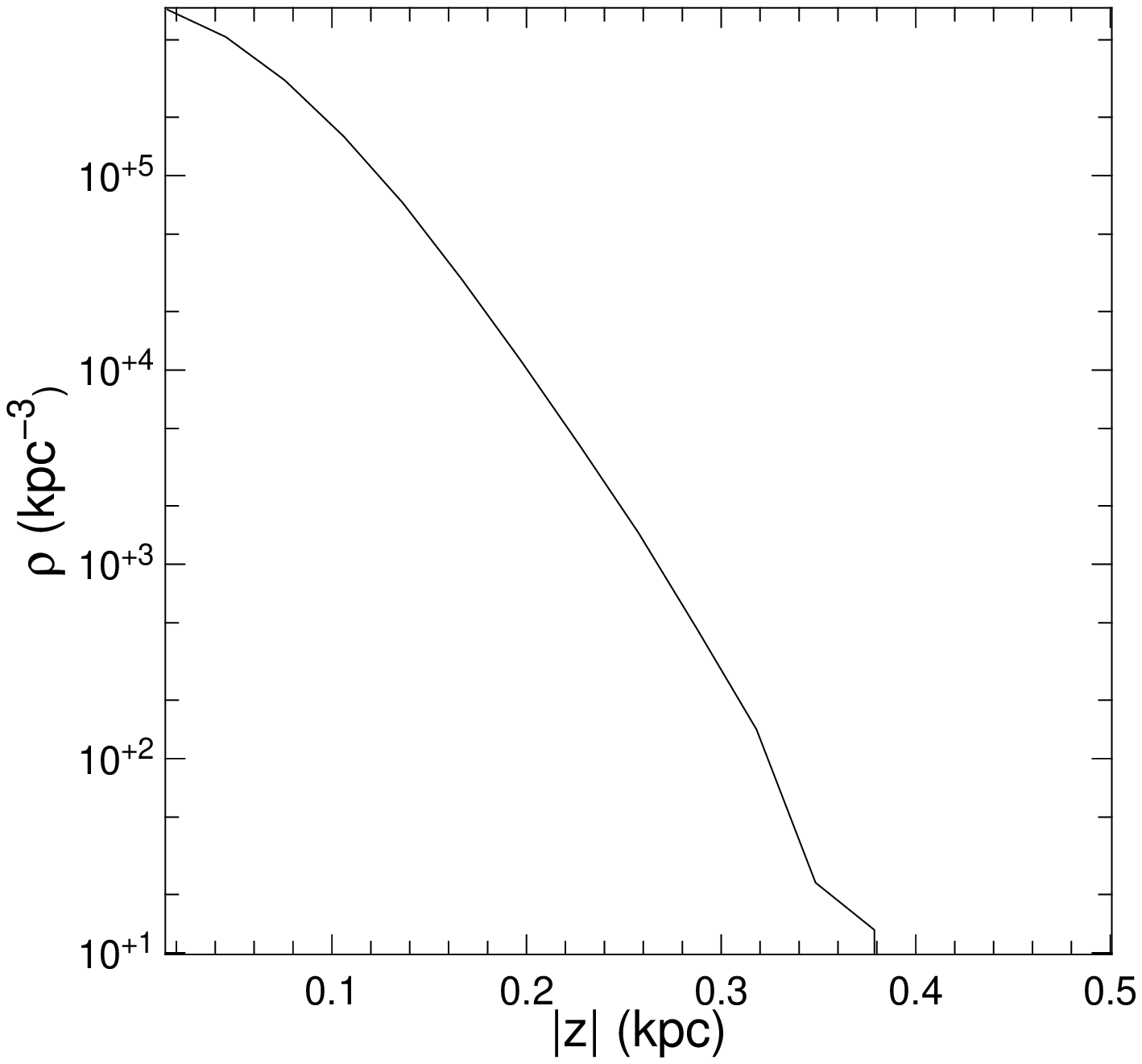}
\caption{Initial conditions. Left panel: Number of stars per ${\rm kpc}^2$ (surface density) within the Galactic plane as a function of $R$. Right panel: Stellar density as a function of $z$ at $R=8$~kpc.}
\label{f:rhoini}
\end{figure}

From  this  distribution  function,  $4\times 10^7$  test  particle  initial
conditions are generated in a 3D polar grid between $R=4$~kpc and $R=15$~kpc
(see  Fig.~\ref{f:rhoini}).  This allows  a  good  resolution  in the  solar
suburb. Before adding the spiral  perturbation, the simulation is run in the
axisymmetric  potential for two  rotations ($\sim  500$~Myr), and  is indeed
stable.

\begin{figure}
\includegraphics[width=7cm]{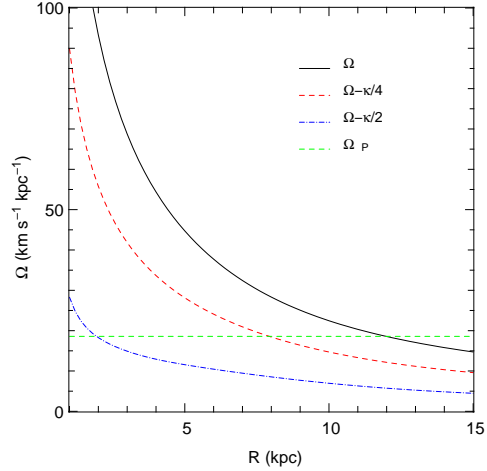}
\caption{Positions   of   the  main   radial   resonances   of  the   spiral
  potential. $\Omega(R  ) = v_c(R)/r$  is the local circular  frequency, and
  $v_c(R )$ is  the circular velocity. The $2:1$ ILR  occurs along the curve
  $\Omega(R  ) -  \kappa/2$, where  $\kappa$ is  the local  radial epicyclic
  frequency.  The inner  $4:1$ IUHR  occurs along  the curve  $\Omega(R  ) -
  \kappa/4$.}
\label{f:Omega}
\end{figure}

\begin{figure}
\includegraphics[width=7cm]{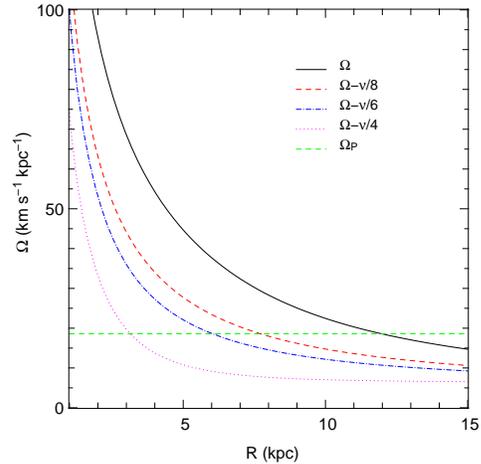}
\caption{Positions of the 4:1, 6:1 and 8:1 vertical resonances. When $\Omega
  - \Omega_P = \nu/n$, where  $\nu$ is the vertical epicyclic frequency, the
  star makes  precisely $n$ vertical oscillations along  one rotation within
  the rotating frame of the spiral.}
\label{f:Omeganu}
\end{figure}

\subsection{Spiral perturbation and orbit integration}

In 3D, we consider a spiral arm perturbation of the Lin-Shu type (Lin \& Shu
1964; see  also Siebert  et al.  2012) with a  sech$^2$ vertical  profile (a
pattern  that can  be supported  by three-dimensional  periodic  orbits, see
e.g. Patsis \& Grosb{\o}l 1996) and a small ($\sim 100$~pc) scale-height:
\begin{equation}
\Phi_{s}(R,\theta,z)=-A \cos\left[m\left( \Omega_P t -
    \theta+\frac{\ln(R)}{\tan p}\right) \right] {\rm sech}^2 \left(\frac{z}{z_0} \right)
\label{spipot}
\end{equation}
%\begin{eqnarray}
%\Phi_{s}(R,\theta,z)&=&-A \cos\left[m\left( \Omega_P t - \theta+\frac{\ln(R)}{\tan p}\right) \right] 
%\nonumber\\&&\times \, {\mathrm sech}^2 \left(\frac{z}{z_0} \right)
%\end{eqnarray}
in which $A$ is the amplitude of the perturbation, $m$ is the spiral pattern
mode ($m=2$ for a 2-armed spiral),  $\Omega_P$ is the pattern speed, $p$ the
pitch angle,  and $z_0$  is the spiral  scale-height. The edge-on  shapes of
orbits  of these  thick spirals  are determined  by the  vertical resonances
existing in the potential.

The parameters of  the spiral potential used in  our simulation are inspired
by the analytic solution found in  Siebert et al. (2012) using the classical
2D Lin-Shu formalism to fit  the radial velocity gradient observed with RAVE
(Siebert  et  al.   2011b).   The  parameters used  here  are  summarized  in
Table~2. The amplitude $A$ which we use corresponds to 1\% of the background
axisymmetric potential at the Solar  radius (3\% of the disc potential). The
positions  of the  main  radial  resonances, i.e.   the  2:1 inner  Lindblad
resonance  (ILR) and 4:1  IUHR, are  illustrated in  Fig.~\ref{f:Omega}. The
presence of the 4:1 IUHR close to the Sun is responsible for the presence of
the Hyades and Sirius moving groups in the local velocity space at the Solar
radius (see  Pomp\'eia et al.   2011), associated to  square-shaped resonant
orbital families in the rotating spiral frame.  Vertical resonances are also
displayed in Fig.~\ref{f:Omeganu}.

Such a spiral perturbation can grow naturally in self-consistent simulations
of isolated discs  without the help of any  external perturber (e.g. Minchev
et al. 2012).  As we are interested hereafter in the  global response of the
thin disc stellar population to  a quasi-static spiral perturbation, we make
sure  to  grow  the  perturbation  adiabatically by  multiplying  the  above
potential perturbation by a growth factor starting at $t\approx 0.5$~Gyr and
finishing  at $t\approx 3.5$~Gyr:  $\epsilon (t)=\frac{1}{2}(\tanh(1.7\times
t-3.4)+1)$.  The integration  of orbits  is performed  using a  fourth order
Runge-Kutta algorithm run on Graphics Processing Units (GPUs). The growth of
this spiral is not  meant to be realistic, as we are  only interested in the
orbital structure of the old thin disk once the perturbation is stable.

\begin{table}
  \caption{Parameters of the spiral potential and location of the main resonances}
  \label{tab:spiral}
  \centering
  \begin{center}
    \begin{tabular}{@{\extracolsep{-5pt}}lc}\hline\hline
      Parameter     & Spiral potential \\ \hline
      $m$ & 2 \\
      $A$ (km$^2$ s$^{-2}$)  & 1000\\
      $p$ (deg)  & -9.9\\
      $z_0(\Kpc)$   & 0.1\\
      $\Omega_P(\km {\rm s}^{-1} \Kpc^{-1})$   & 18.6\\
      $R_{\rm ILR}(\Kpc)$ & 1.94\\
      $R_{\rm IUHR}(\Kpc)$ & 7.92\\
      $\RCR(\Kpc)$   &  11.97\\\hline
    \end{tabular}
  \end{center}
\end{table}

\begin{figure*}
\includegraphics[width=6cm]{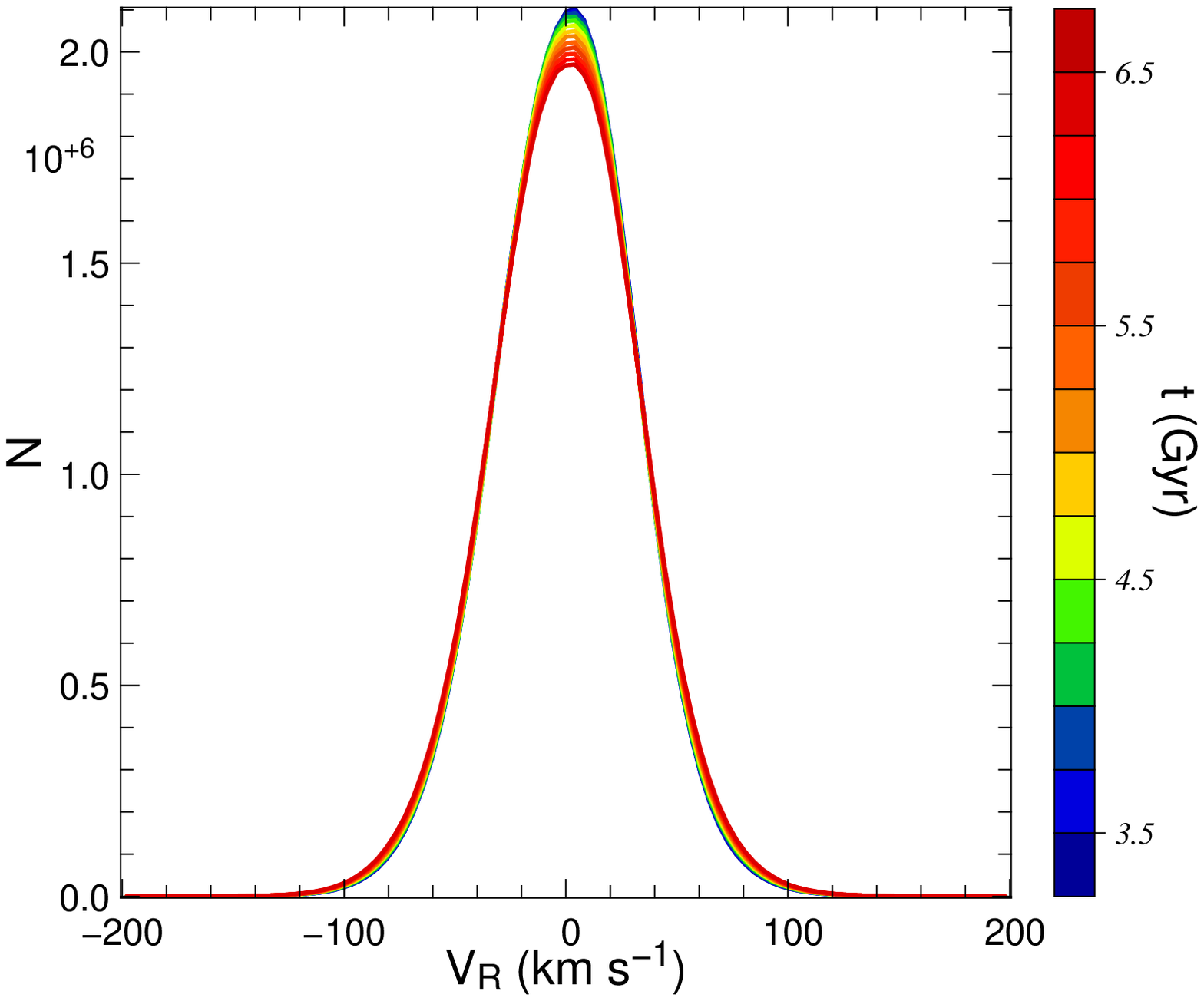}
\includegraphics[width=6cm]{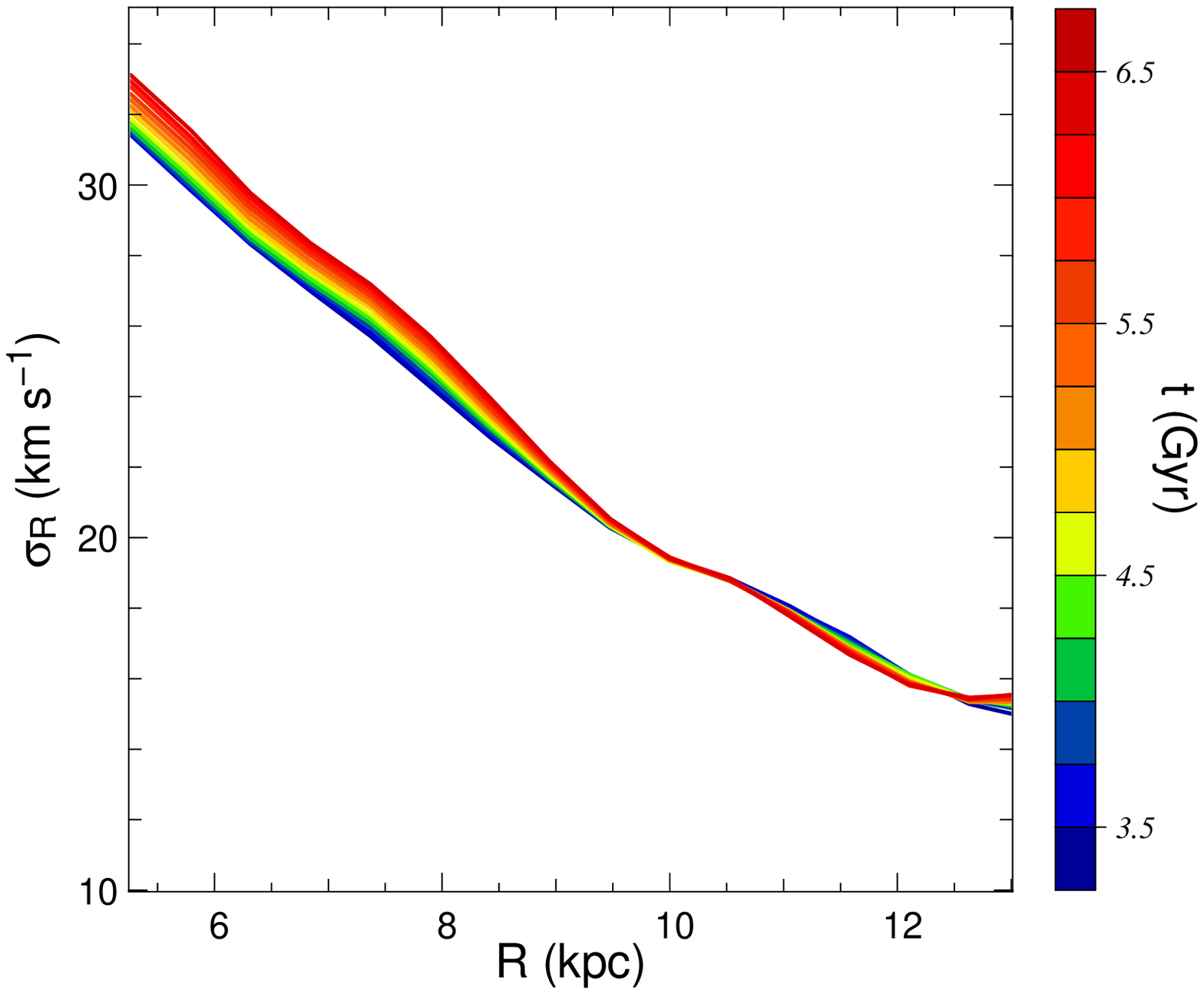}
\caption{Left  panel: Histogram  of  galactocentric radial  velocities as  a
  function of time. Right panel: time evolution of the $\sigma_R(R)$ profile
  averaged over  all azimuths. The  colour-scale indicates the  time-steps in
  Gyr.}
\label{f:sigmar}
\end{figure*}

\begin{figure*}
\includegraphics[width=6cm]{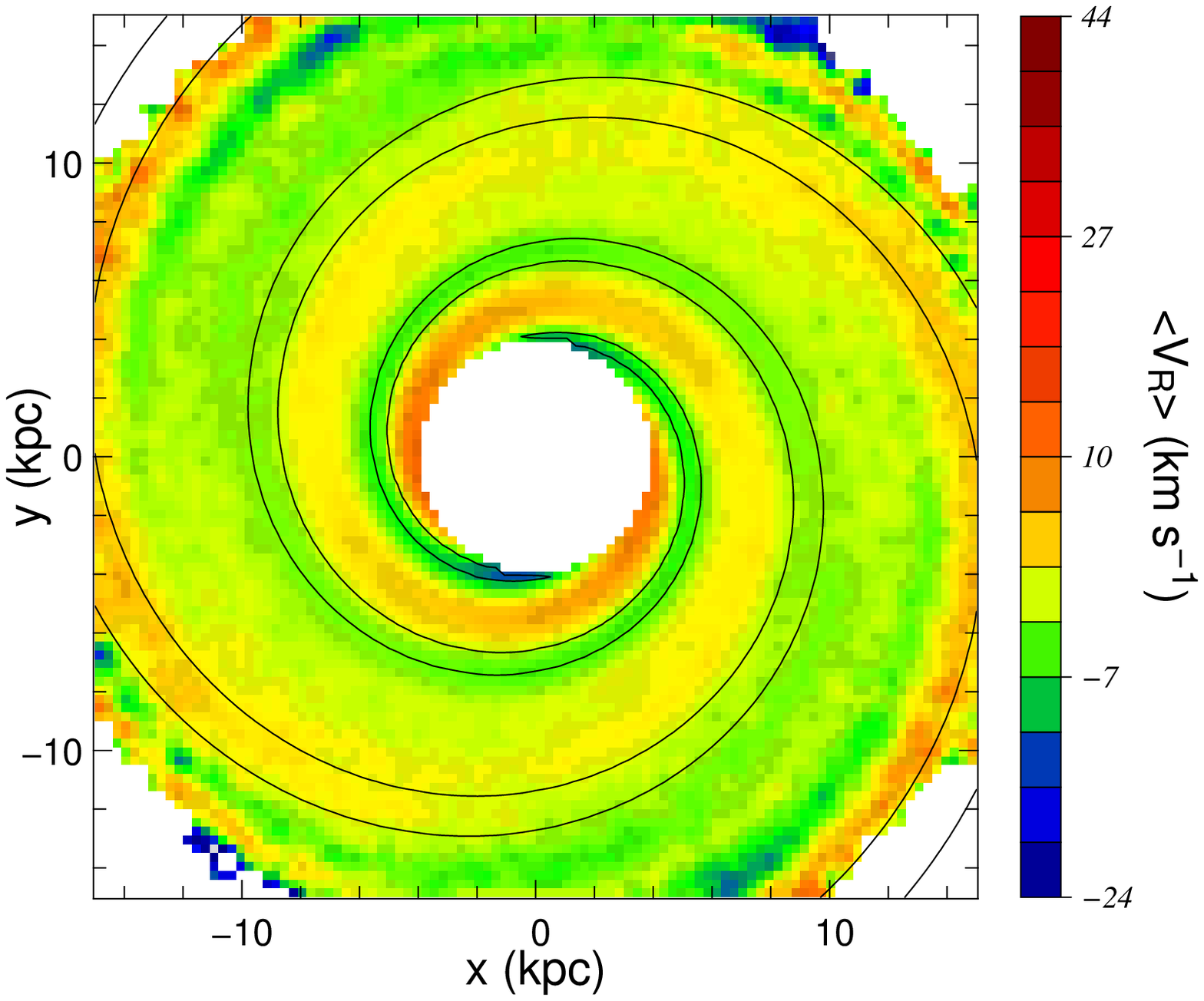}
\includegraphics[width=6cm]{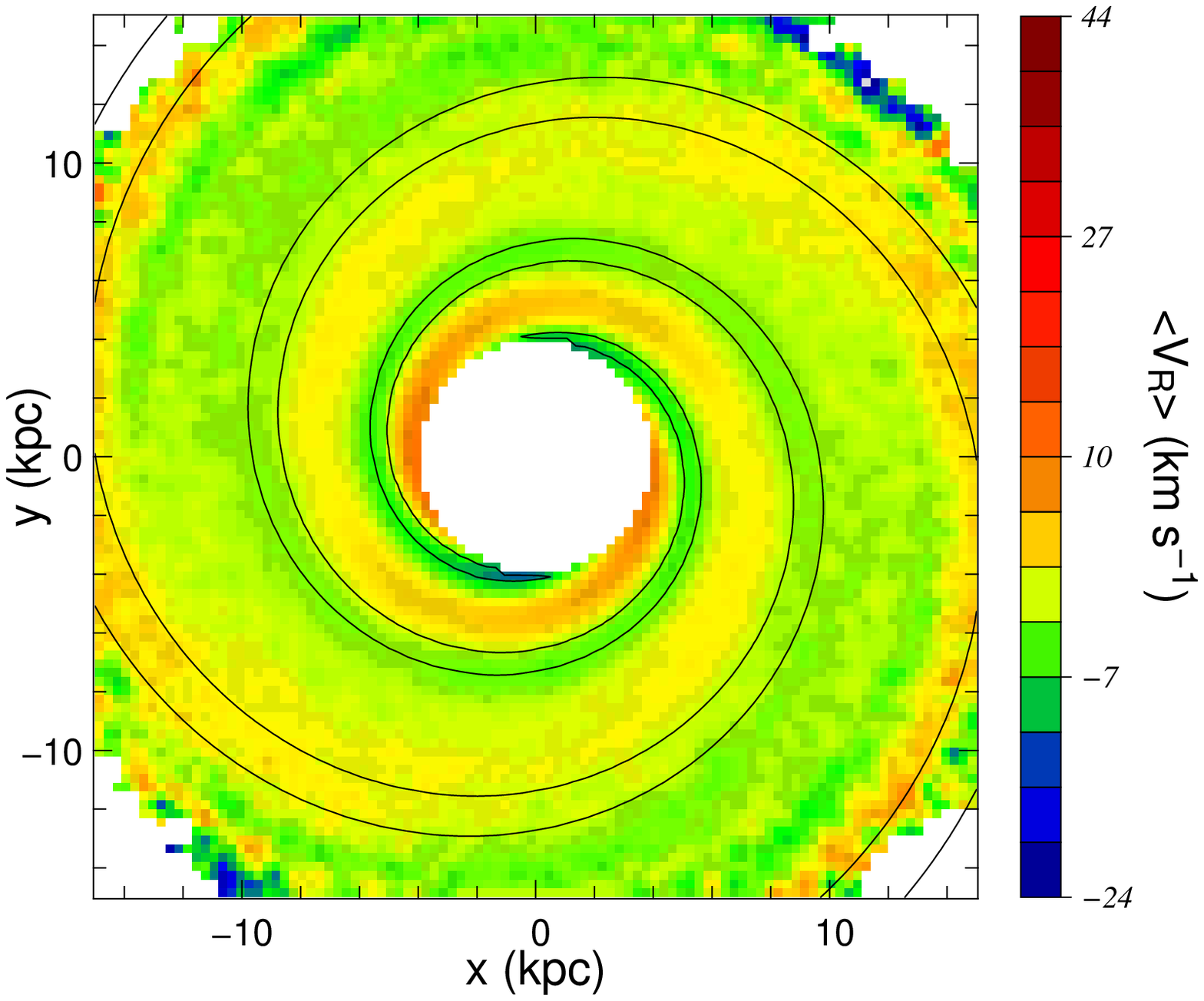}
\includegraphics[width=6cm]{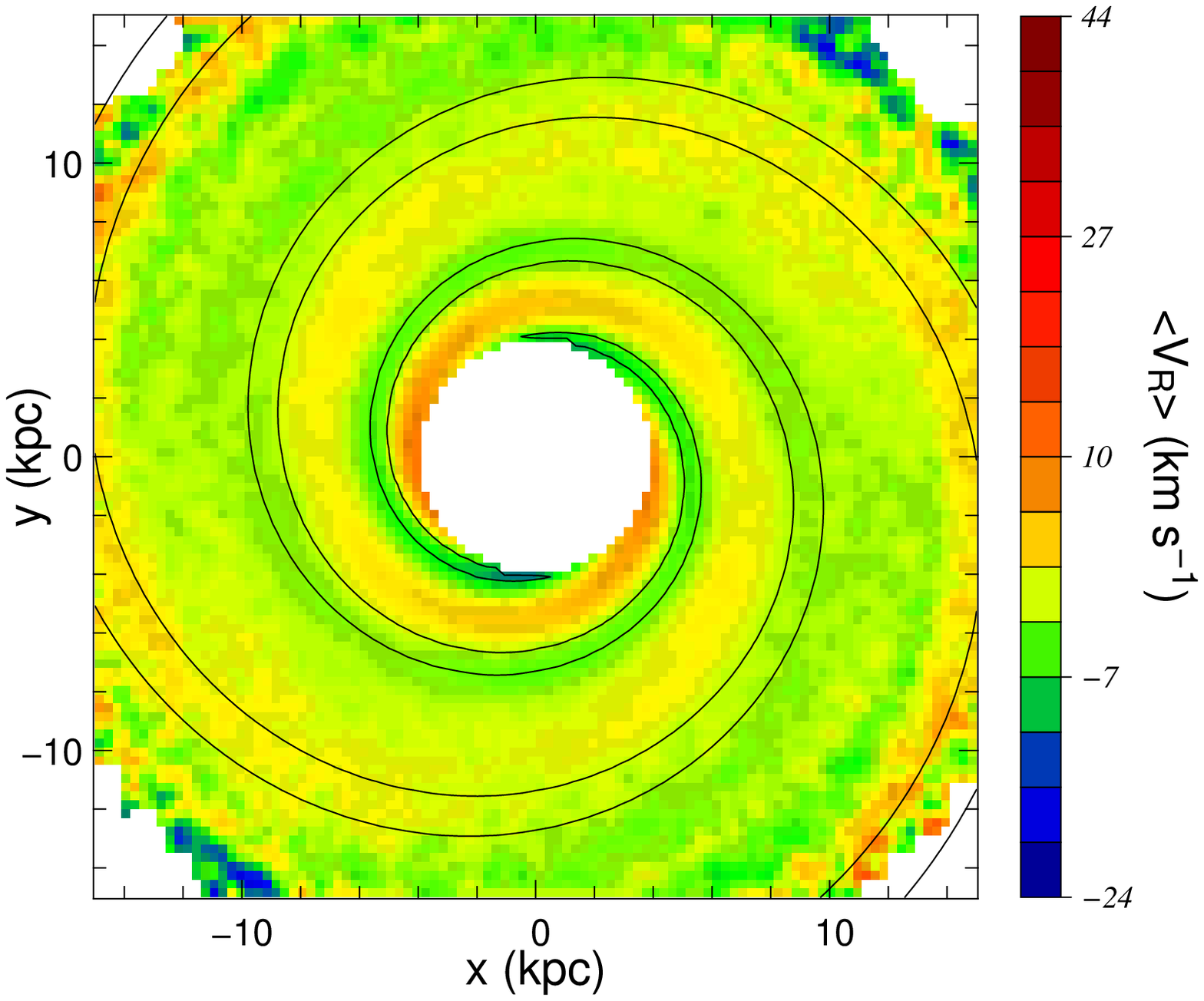}
\includegraphics[width=6cm]{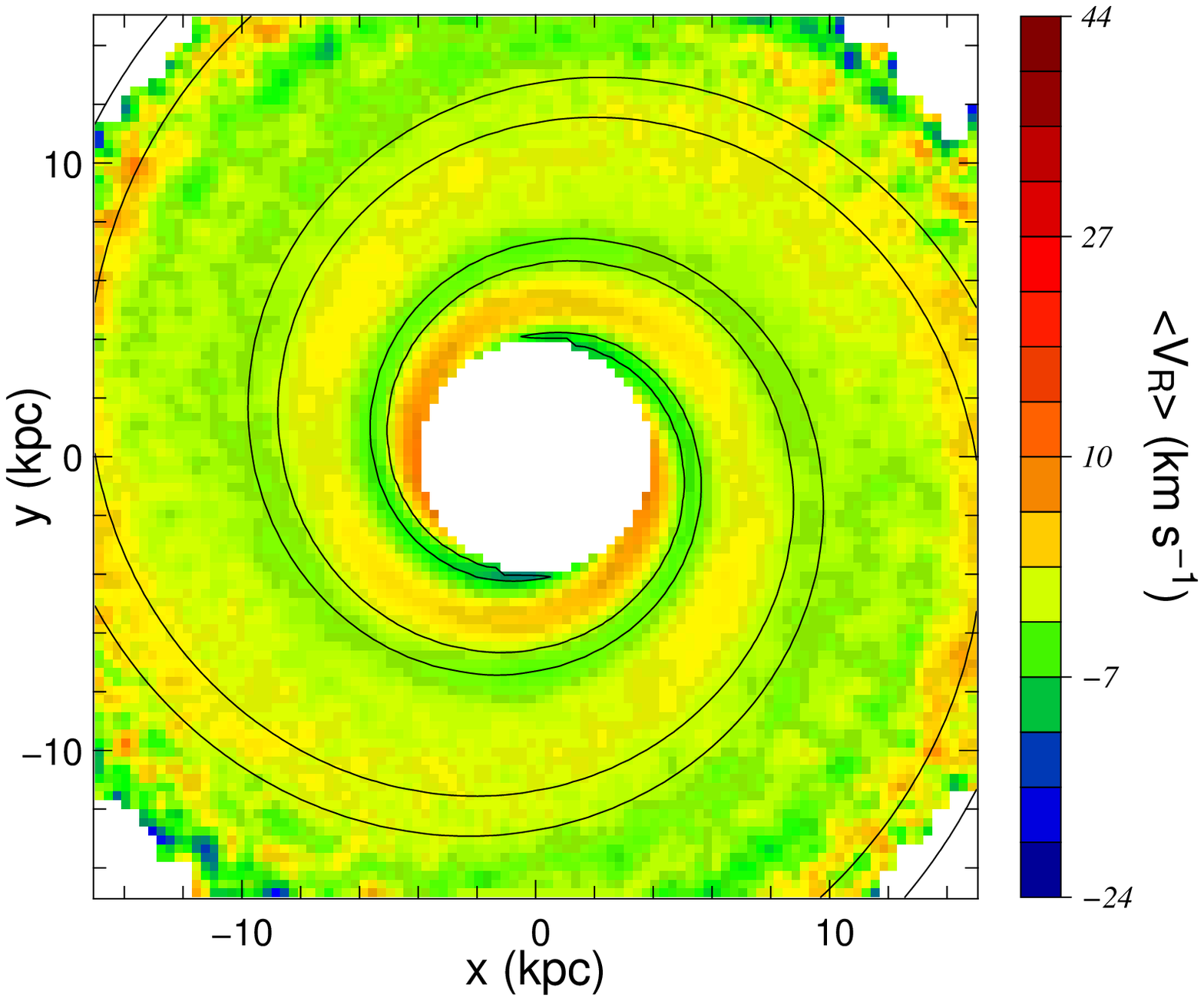}
\caption{Top-left  panel: Mean galactocentric  radial velocity  $\langle v_R
  \rangle$ as  a function of position  in the Galactic plane  soon after the
  adiabatic growth  of the spiral  ($t=4 \,$Gyr). Isocontours of  the spiral
  potential are  overplotted, corresponding  to 80\% of  the minimum  of the
  perturber  potential, and  thus  delimiting the  region  where the  spiral
  potential is between 80\% and 100\% of its minimum (or maximum in absolute
  value). Top right: Same at  $t=5 \,$Gyr. Everything is plotted here within
  the rotating  frame of the spiral, so  that the spiral does  not move from
  one snapshot to the other. Bottom left: Same at $t=6 \,$Gyr. Bottom right:
  Same at the final time-step $t=6.5 \,$Gyr.}
\label{f:cartevr}
\end{figure*}

\begin{figure}
\includegraphics[width=8cm]{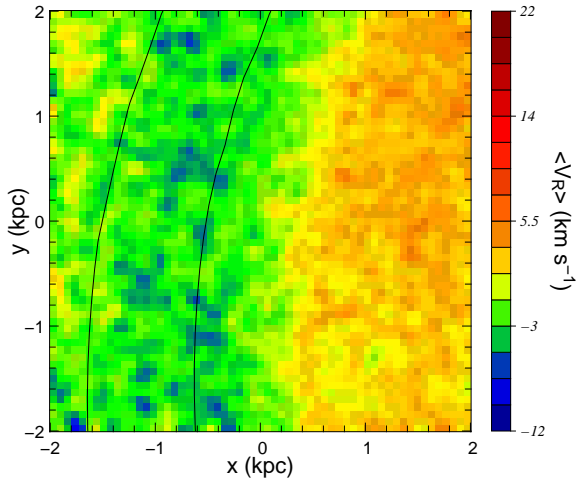}
\caption{Galactocentric radial  velocities in the solar  suburb, centered at
  $(R,\theta)=(8\, {\rm kpc},  26^\circ)$ at $t=4 \,$Gyr. On  this plot, the
  Sun is centered on $(x,y)=(0,0)$,  positive $x$ indicates the direction of
  the Galactic centre,  and positive $y$ the direction  of galactic rotation
  (as well  as the  sense of  rotation of the  spiral pattern).   The spiral
  potential  contours overplotted  (same as  on  Fig.~\ref{f:cartevr}, delimiting the  region  where the  spiral
  potential is between 80\% and 100\% of its absolute maximum) would
  correspond  to  the  location of  the  Perseus  spiral  arm in  the  outer
  Galaxy. This Figure can be  qualitatively compared to Fig.~4 of Siebert et
  al. (2011b) and Fig.~3 of Siebert et al. (2012).}
\label{f:RAVE}
\end{figure}

\section{Results}

\subsection{Radial velocity flow}

The histogram of individual galactocentric radial velocities, $v_R$, as well
as  the time-evolution of  the radial  velocity dispersion  profile starting
from $t=3.5 \,$Gyr  (once the steady spiral pattern  is settled) are plotted
on Fig.~\ref{f:sigmar}.   It can be  seen that these are  reasonably stable,
and  that the mean  radial motion  of stars  is very  close to  zero (albeit
slightly  positive).   Our  test  population  is  thus   almost  in  perfect
equilibrium.

However, due to the presence  of spiral arms, the mean galactocentric radial
velocity $\langle v_R  \rangle$ of our test population  is non-zero at given
positions  within the  frame of  the spiral  arms. The  map of  $\langle v_R
\rangle$  as   a  function   of  position  in   the  plane  is   plotted  on
Fig.~\ref{f:cartevr},  for  different time-steps  (4~Gyr,  5~Gyr, 6~Gyr  and
6.5~Gyr). Within the rotating frame  of the spiral pattern, the locations of
these non-zero mean radial velocities  are stable over time: this means that
the response to the spiral perturbation is stable, even though the amplitude
of  the  non-zero velocities  might  slightly  decrease  with time.   Within
corotation, the mean $\langle v_R \rangle$ is negative within the arms (mean
radial  motion towards  the  Galactic centre)  and  positive (radial  motion
towards the  anticentre) between the arms.  Outside  corotation, the pattern
is reversed. This is exactly what  is expected from the Lin-Shu density wave
theory (see, e.g., Eq.~3 in Siebert et al. 2012).

If we place  the Sun at $(R,\theta)=(8\, {\rm kpc},  26^\circ)$ in the frame
of the spiral,  we can plot the expected radial velocity  field in the Solar
suburb (Fig.~\ref{f:RAVE}).  We see  that the galactocentric radial velocity
is  positive in  the inner  Galaxy, as  observed by  Siebert et  al. (2011b),
because the  inner Galaxy  in the local  suburb corresponds to  an inter-arm
region located  within the corotation  of the spiral.   Observations towards
the outer arm (which should correspond  to the Perseus arm in the Milky Way)
should reveal negative galactocentric radial velocities.

\begin{figure*}
\includegraphics[width=6cm]{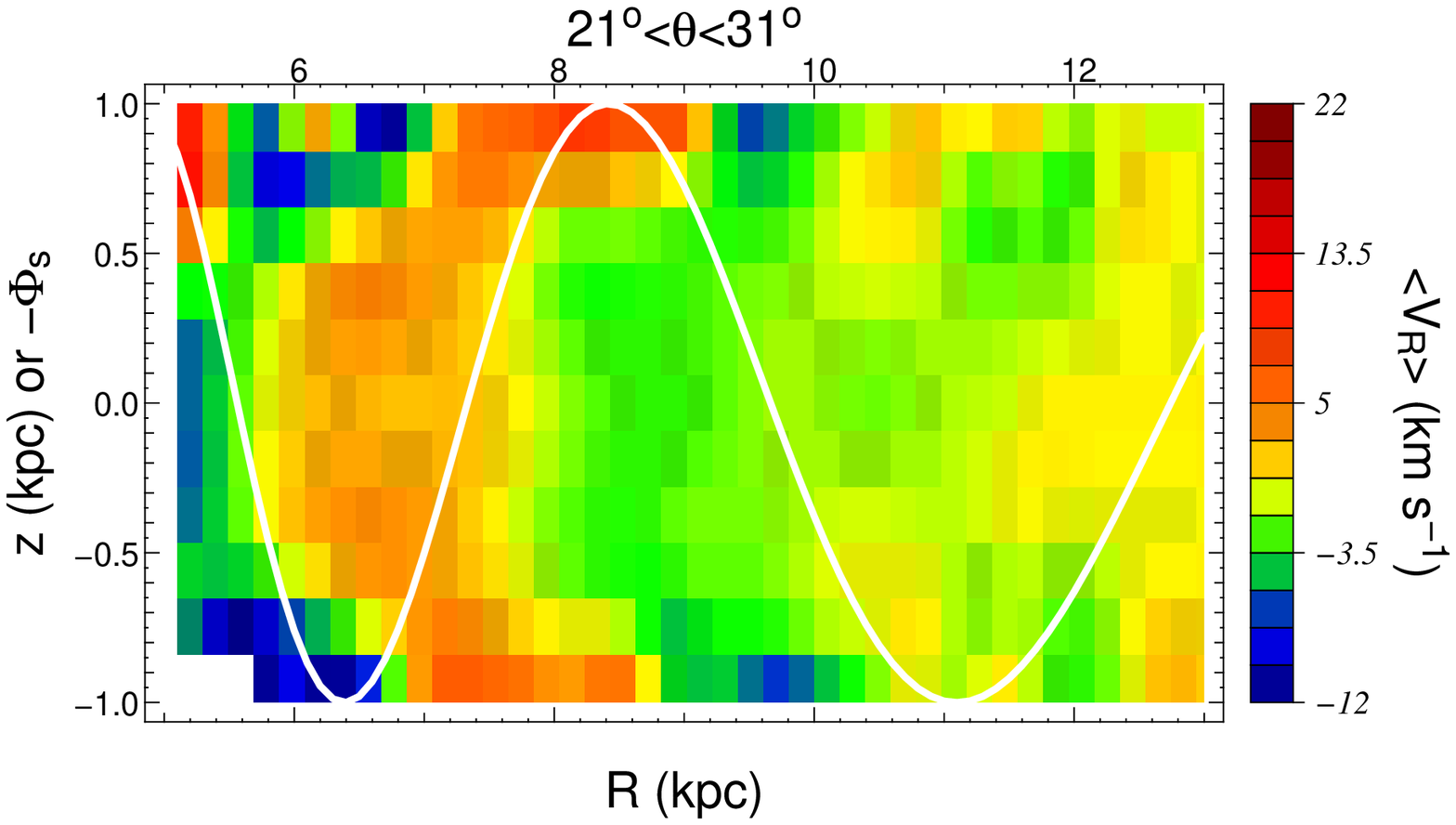}
\includegraphics[width=6cm]{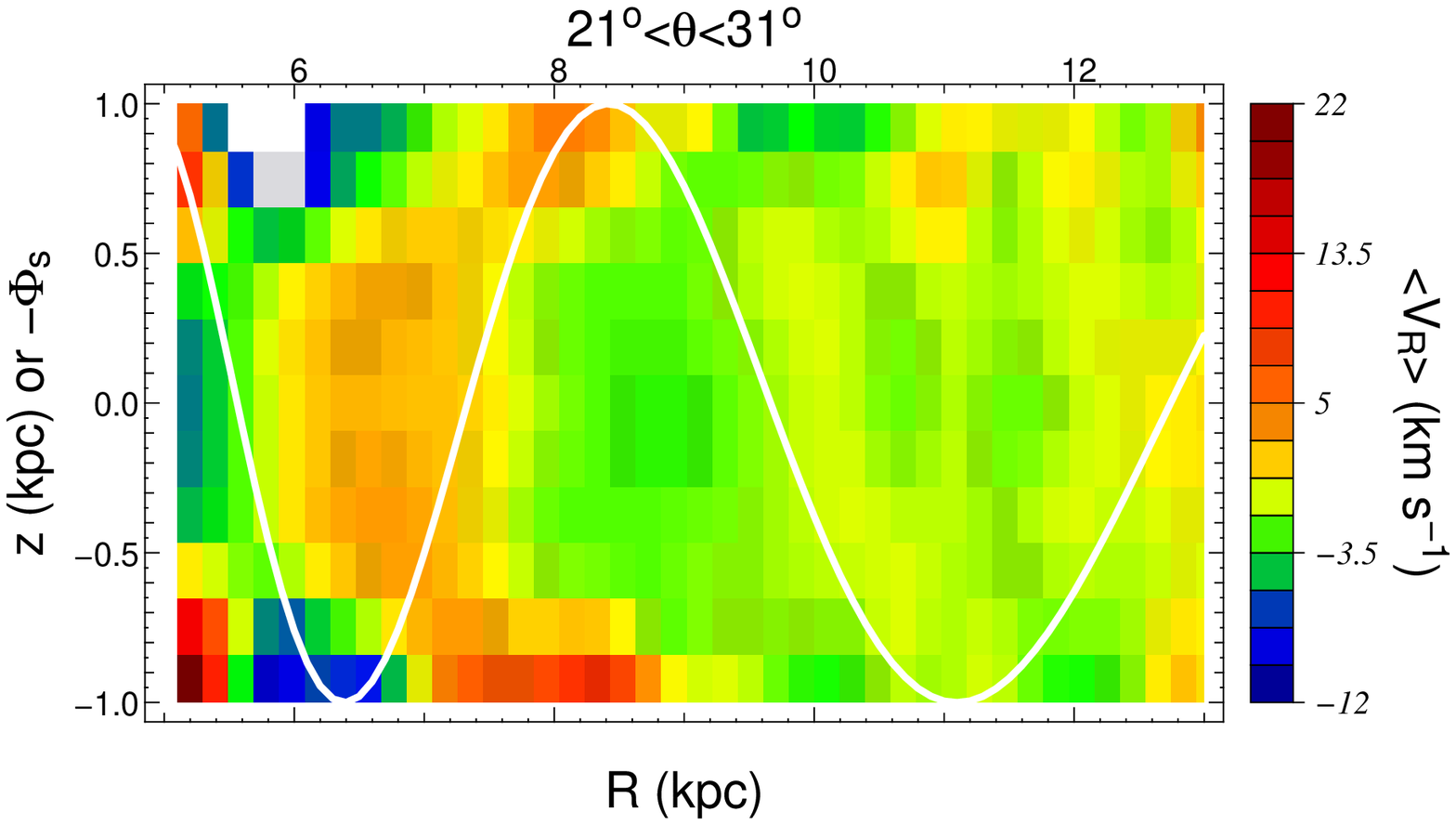}
\includegraphics[width=6cm]{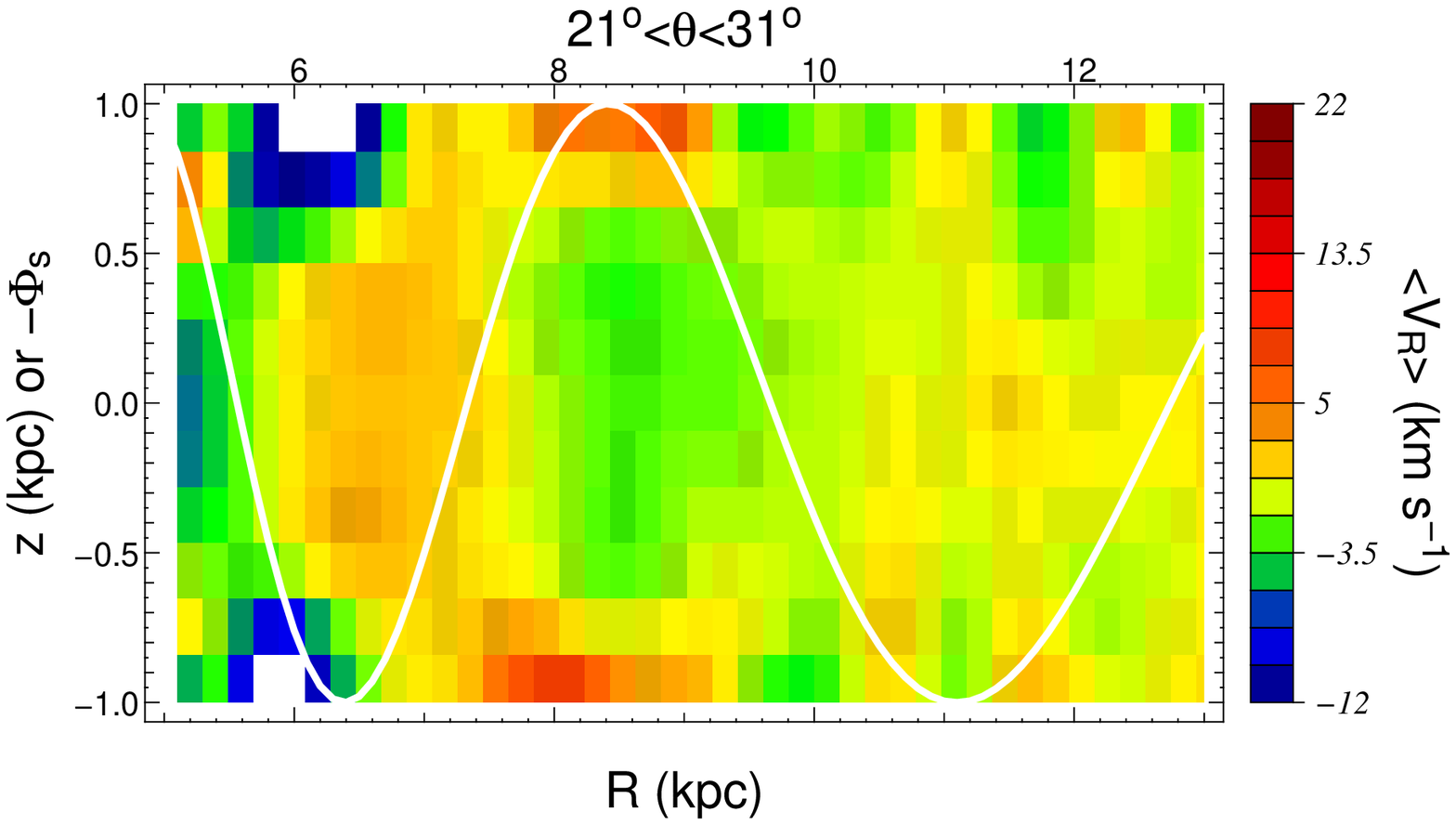}
\includegraphics[width=6cm]{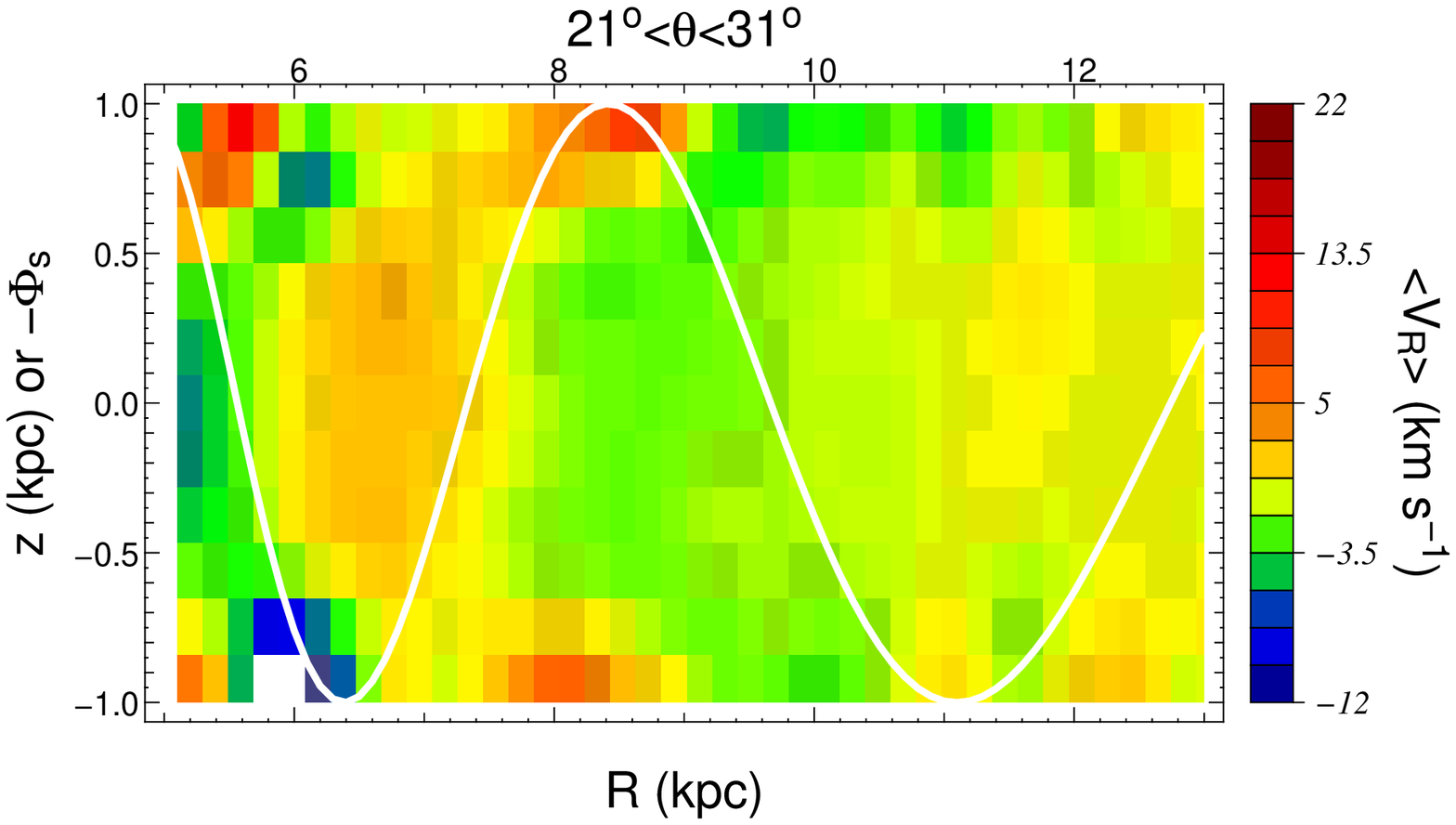}
\caption{Top left panel: Mean galactocentric  radial velocity at $t=4 \,$Gyr
  in the meridional $(R,z)$-plane for $21^\circ<\theta<31^\circ$ (within the
  frame  of the spiral).  The white  line indicates  the location  of spiral
  arms, in  terms of overdensities and underdensities  generating the spiral
  potential (normalized $-\Phi_s$, i.e. spiral arms  are located at the peaks).  Top  right: Same at $t=5
  \,$Gyr. Bottom left: Same at $t=6  \,$Gyr. Bottom right: Same at the final
  time-step $t=6.5 \,$Gyr.}
\label{f:cartevrRZ}
\end{figure*}

\begin{figure*}
\includegraphics[width=6cm]{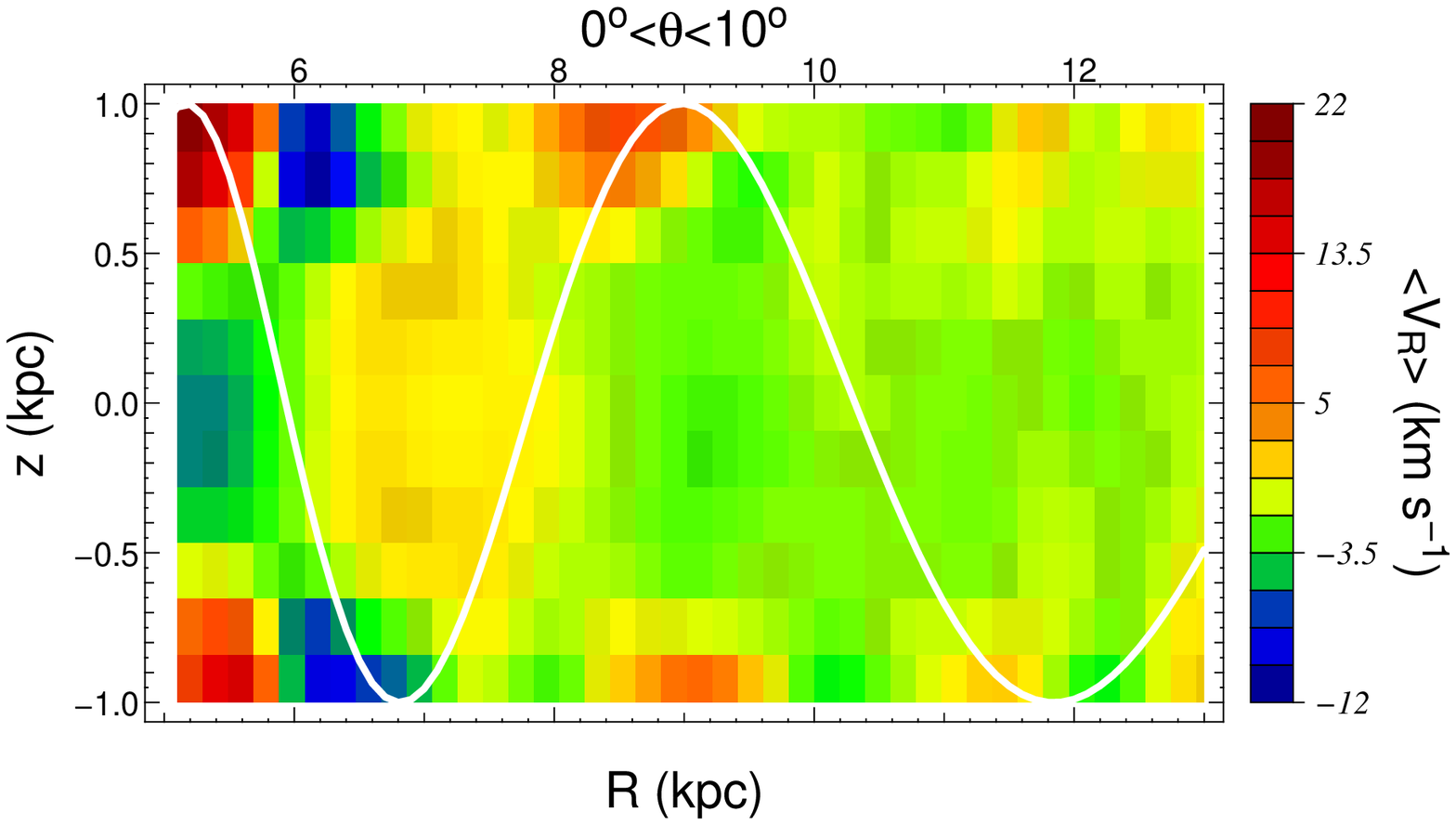}
\includegraphics[width=6cm]{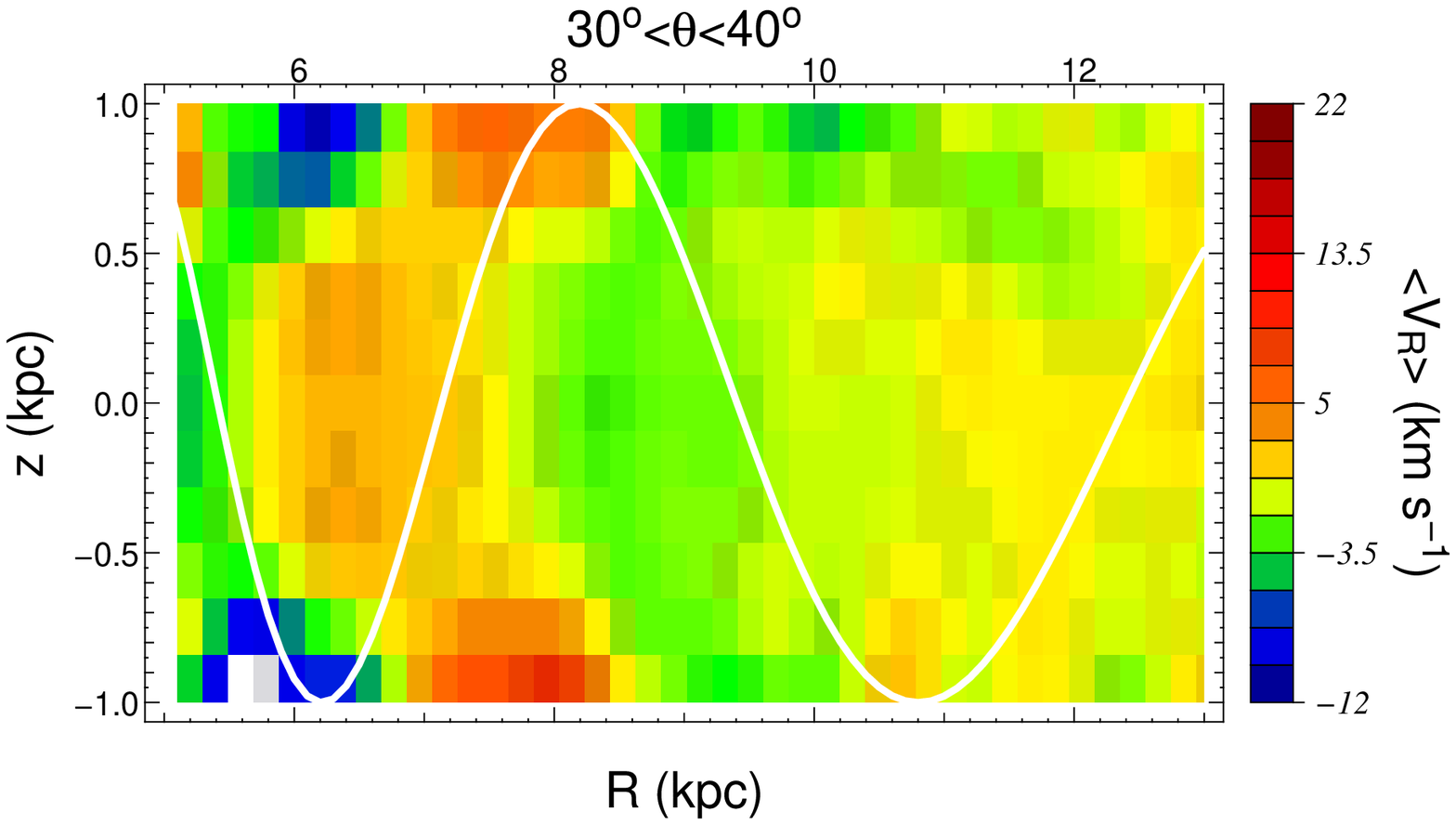}
\includegraphics[width=6cm]{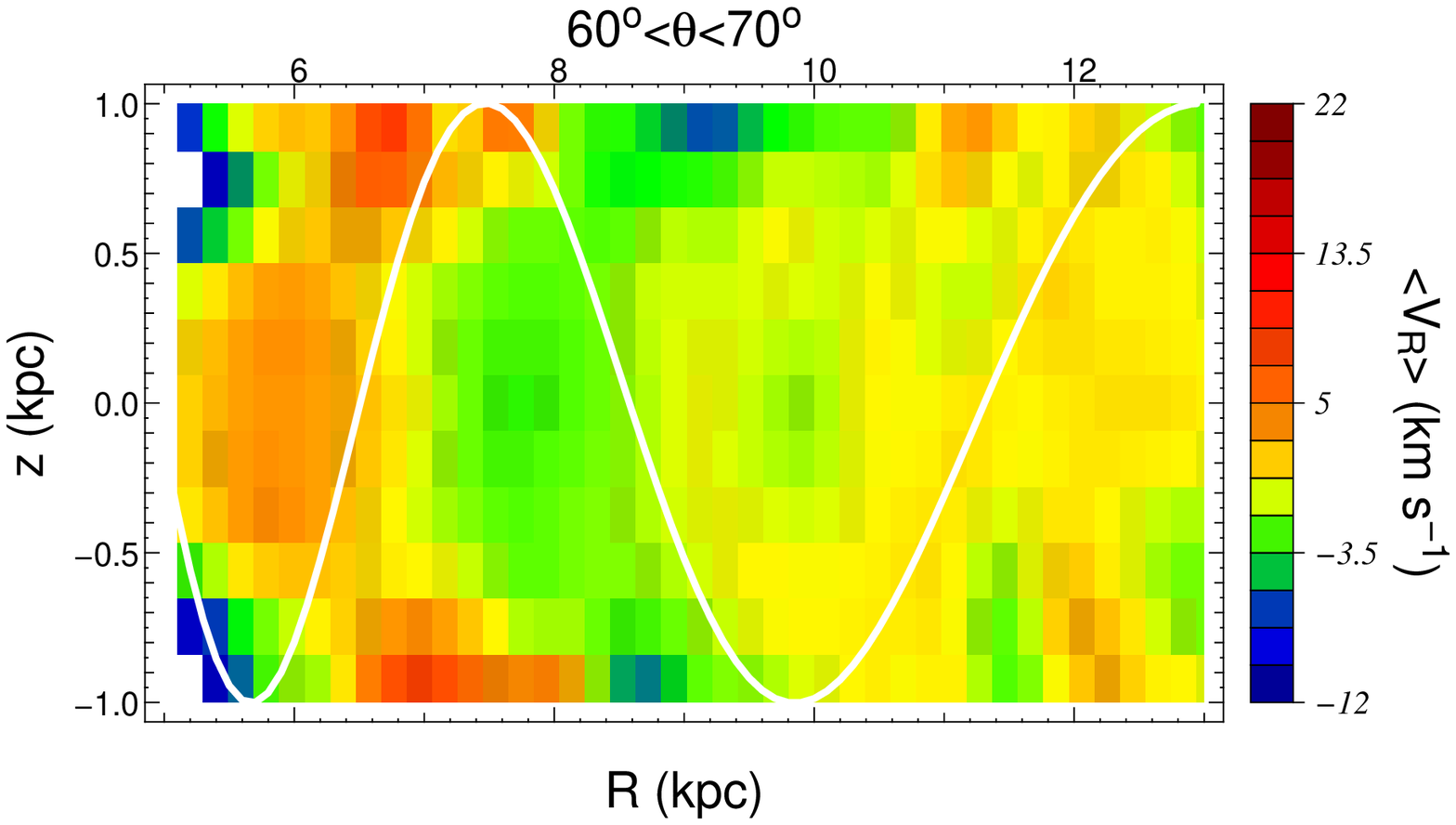}
\includegraphics[width=6cm]{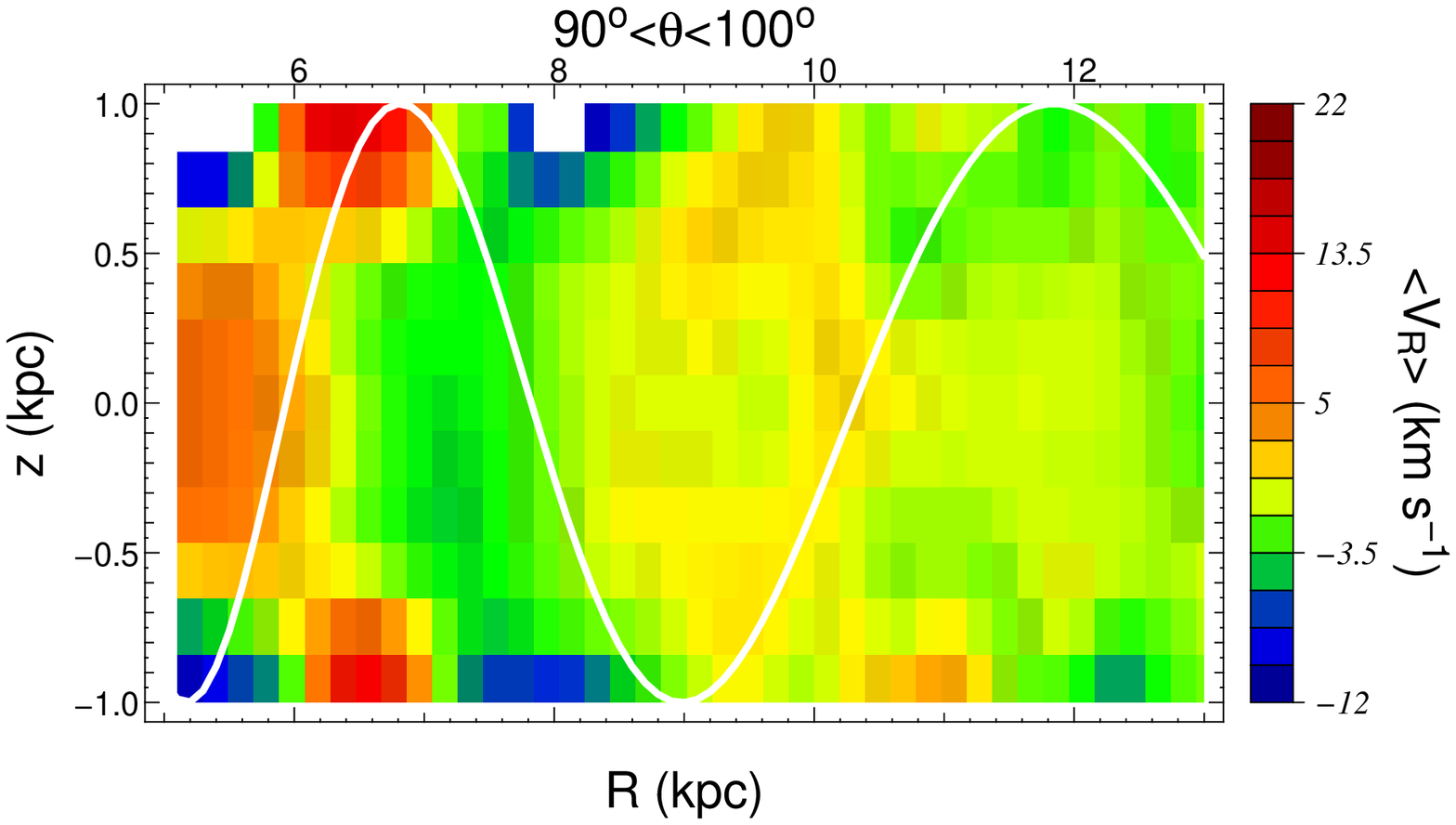}
\includegraphics[width=6cm]{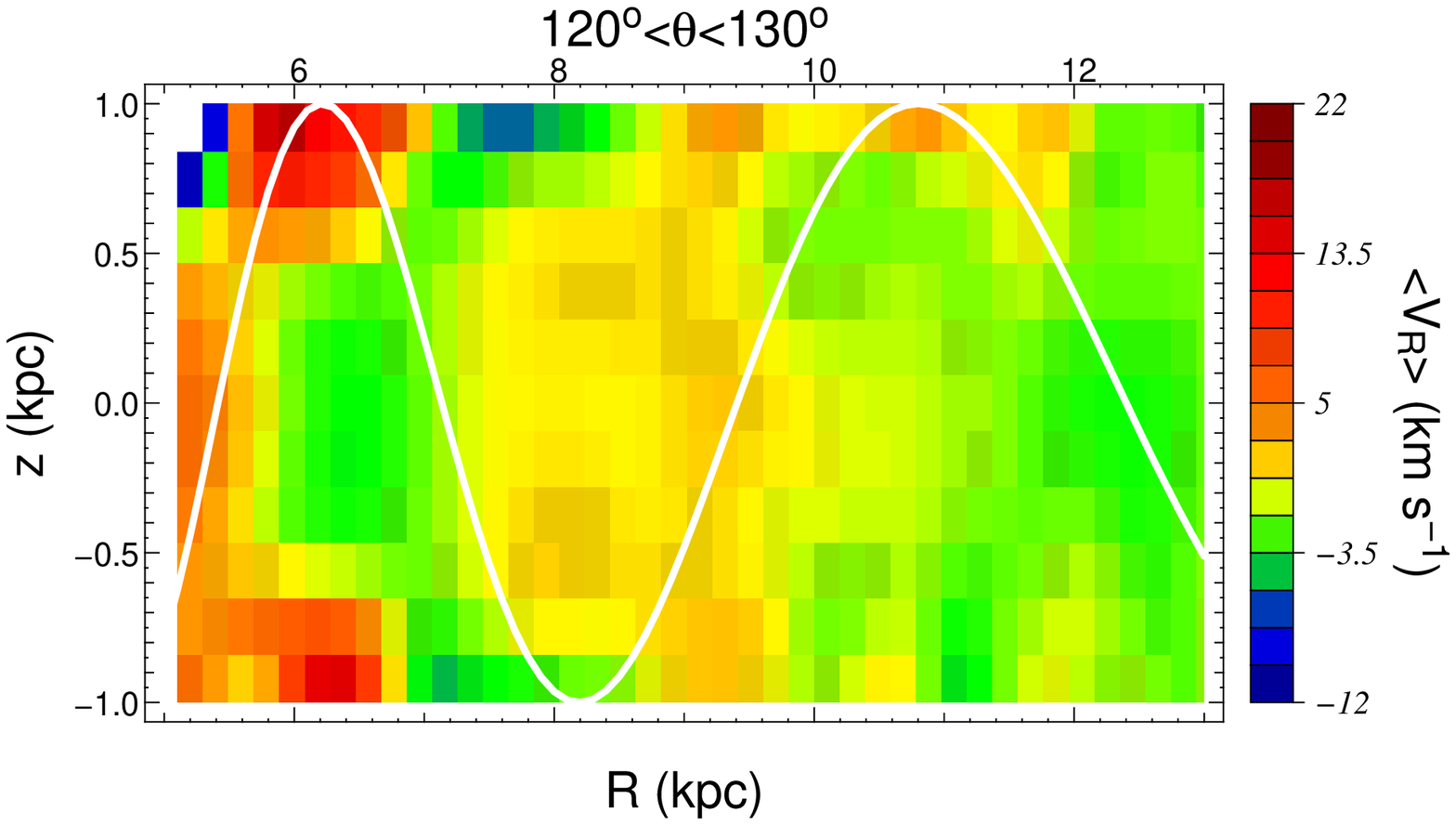}
\includegraphics[width=6cm]{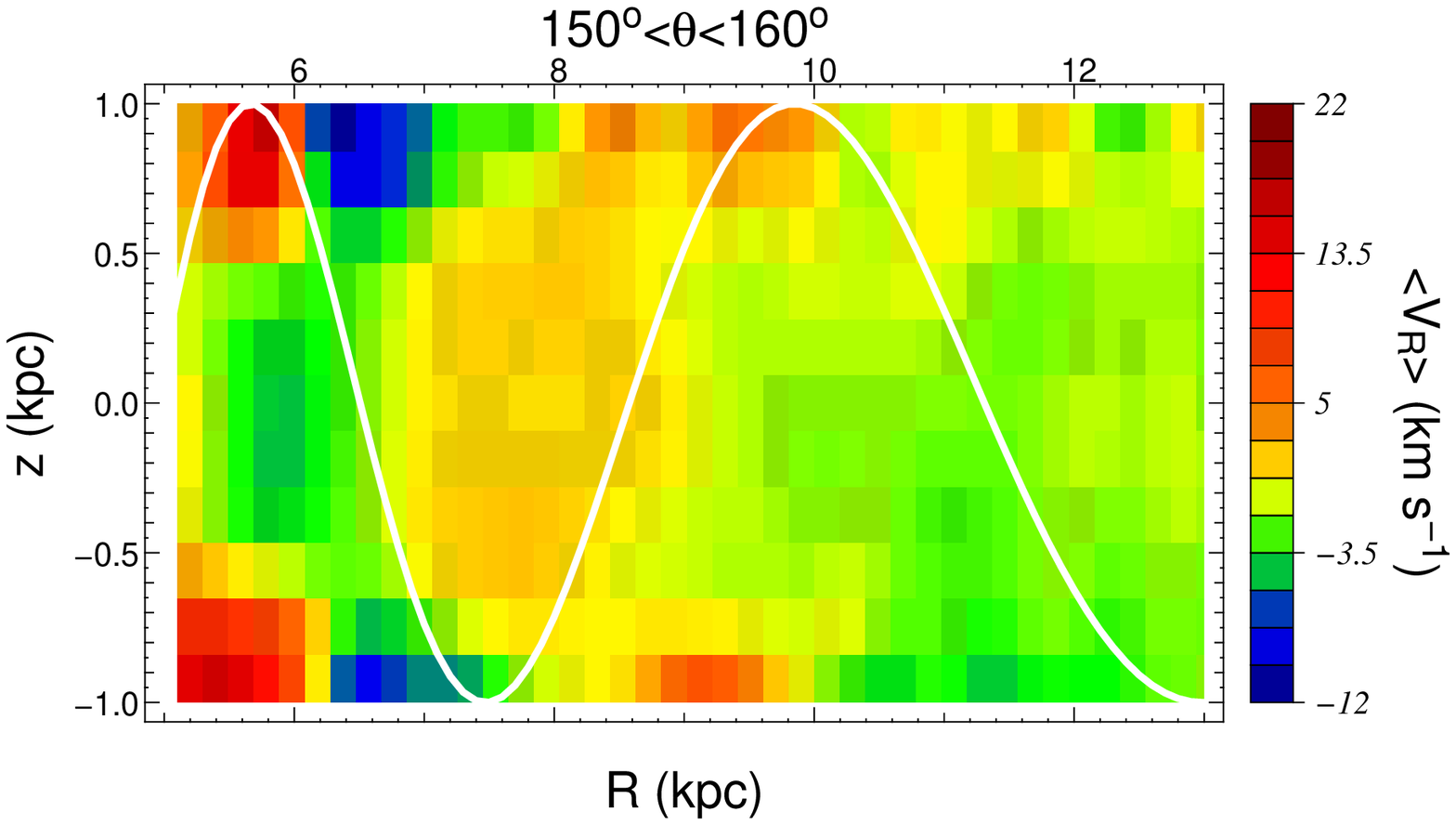}
\caption{Same as Fig.~\ref{f:cartevrRZ}, but for six different azimuths at a
  fixed time ($t=6 \,$Gyr).}
\label{f:cartevrRZ_azimuth}
\end{figure*}

An important aspect of the present study is the behaviour of the response to
a spiral perturbation away from  the Galactic plane. The spiral perturbation
of the potential is very thin in our  model ($z_0 = 100 \,$pc) but as we can
see  on Figs.~\ref{f:cartevrRZ}  and  \ref{f:cartevrRZ_azimuth}, the  radial
velocity flow is not varying much as  a function of $z$ up to five times the
scale-height of the spiral perturber.  This justifies the assumption made in
Siebert et  al. (2012) that the flow  observed at $\sim 500  \,$pc above the
plane was representative  of what was happening in  the plane. Nevertheless,
above these  heights, the trend  seems to be  reversed, probably due  to the
higher  eccentrities of  stars,  corresponding to  different guiding  radii.
This  could potentially  provide a  useful observational  constraint  on the
scale height  of the spiral potential,  a test that could  be conducted with
the forthcoming surveys.

\subsection{Non-zero mean vertical motions}

\begin{figure*}
\includegraphics[width=6cm]{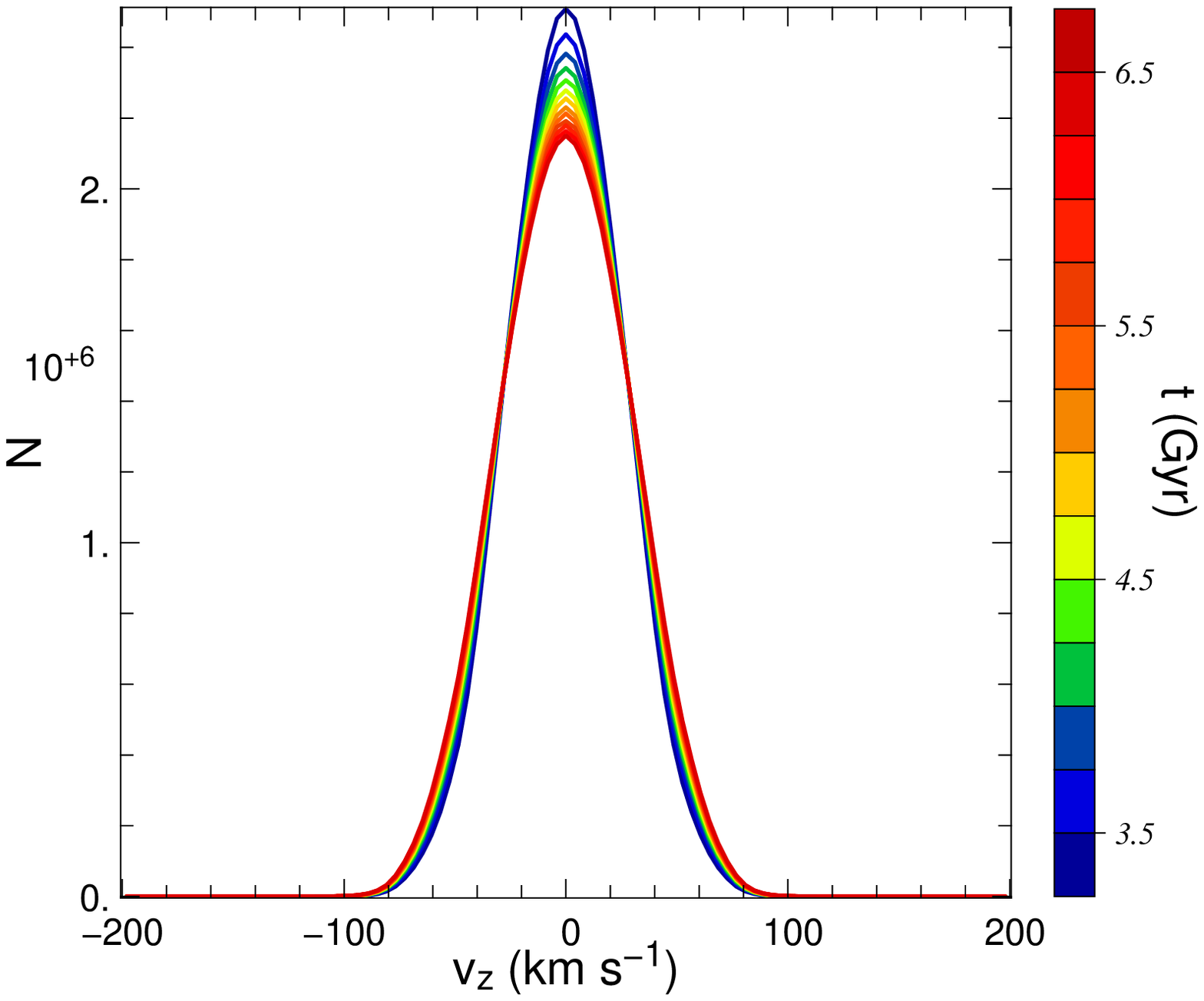}
\includegraphics[width=6cm]{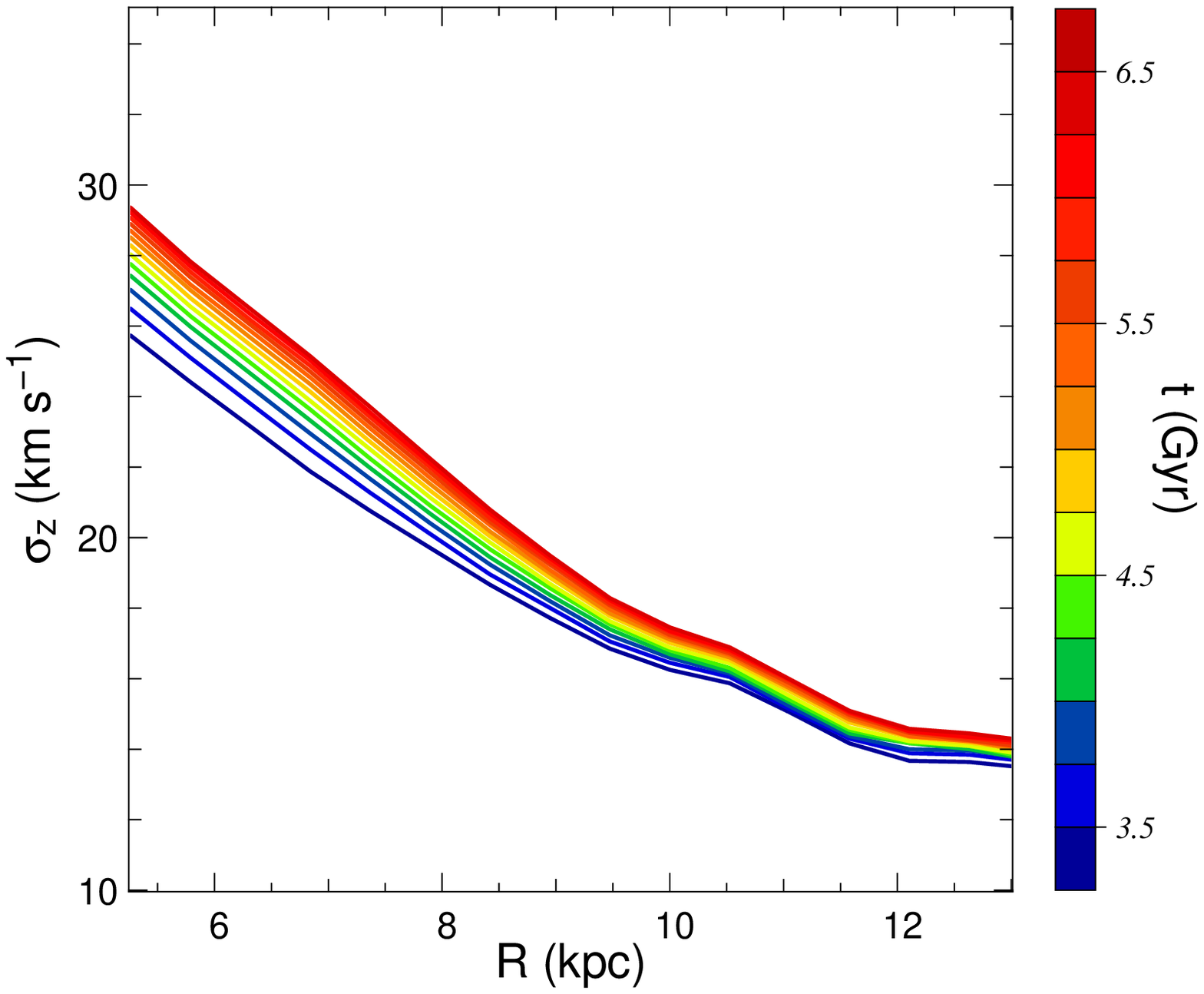}
\caption{Left panel: Histogram of vertical  velocities as a function of time
  . Right panel:  time evolution of the $\sigma_z(R)$  profile averaged over
  all azimuths. }
\label{f:sigmaz}
\end{figure*}

\begin{figure*}
\includegraphics[width=6cm]{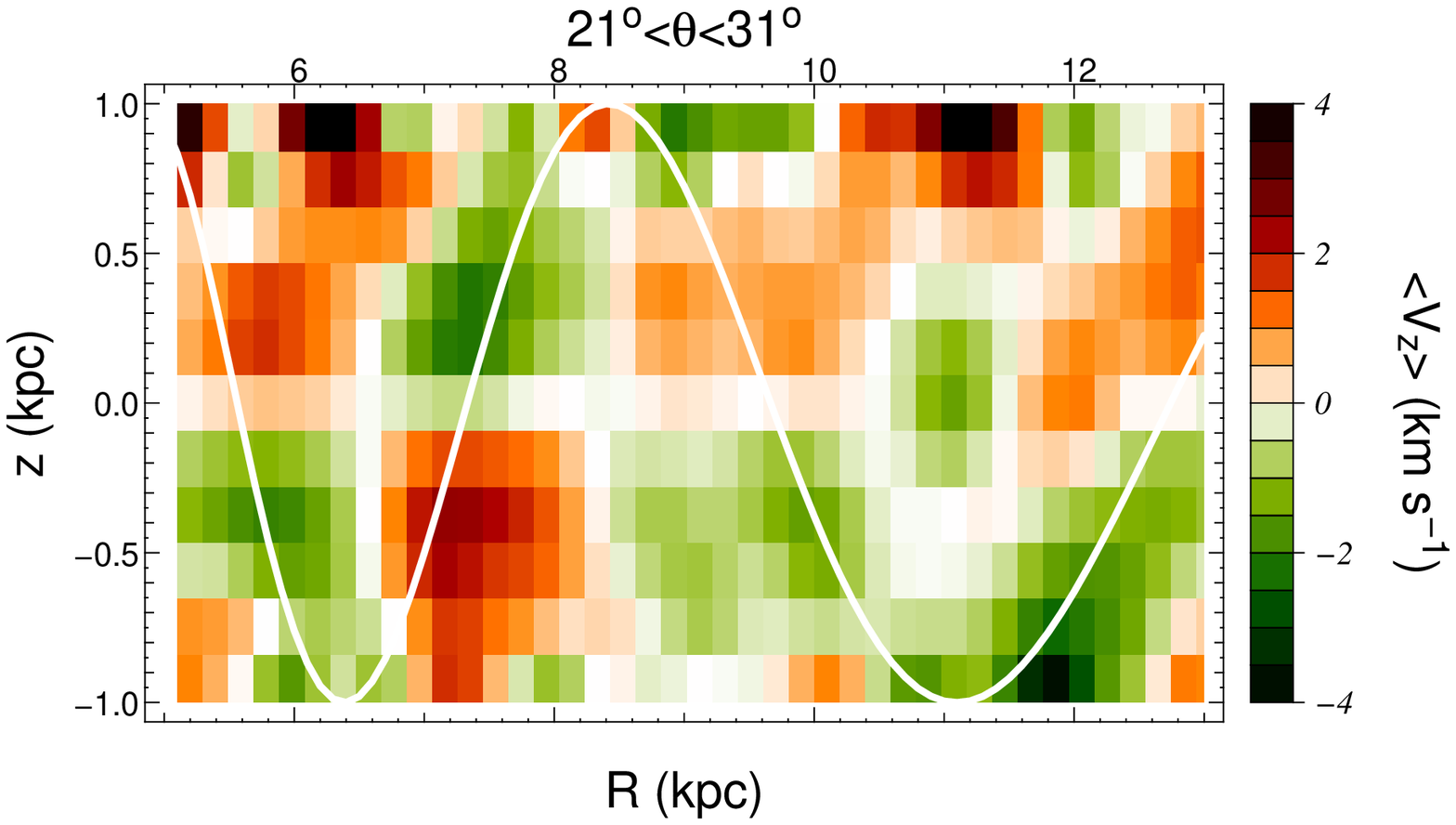}
\includegraphics[width=6cm]{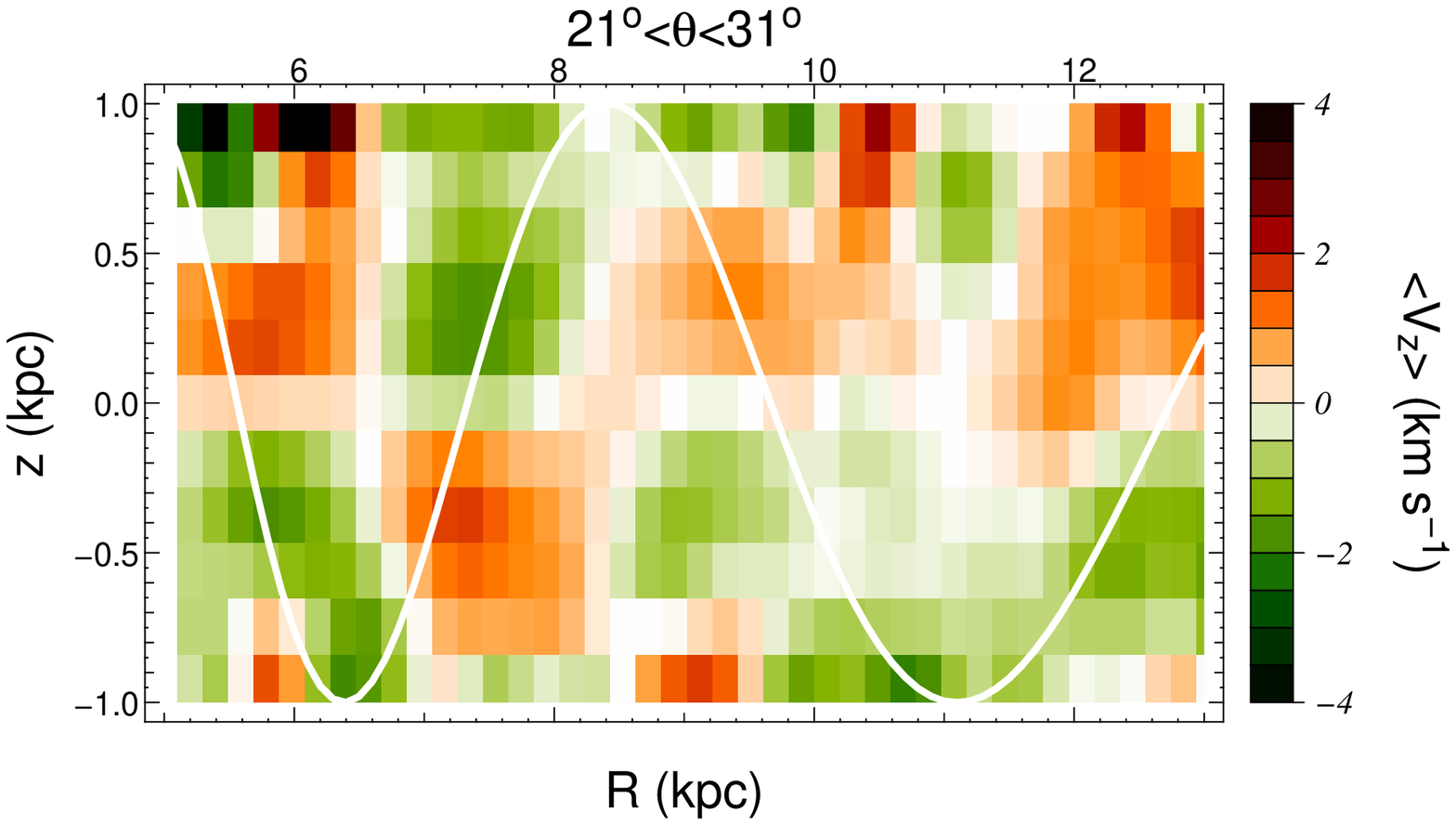}
\includegraphics[width=6cm]{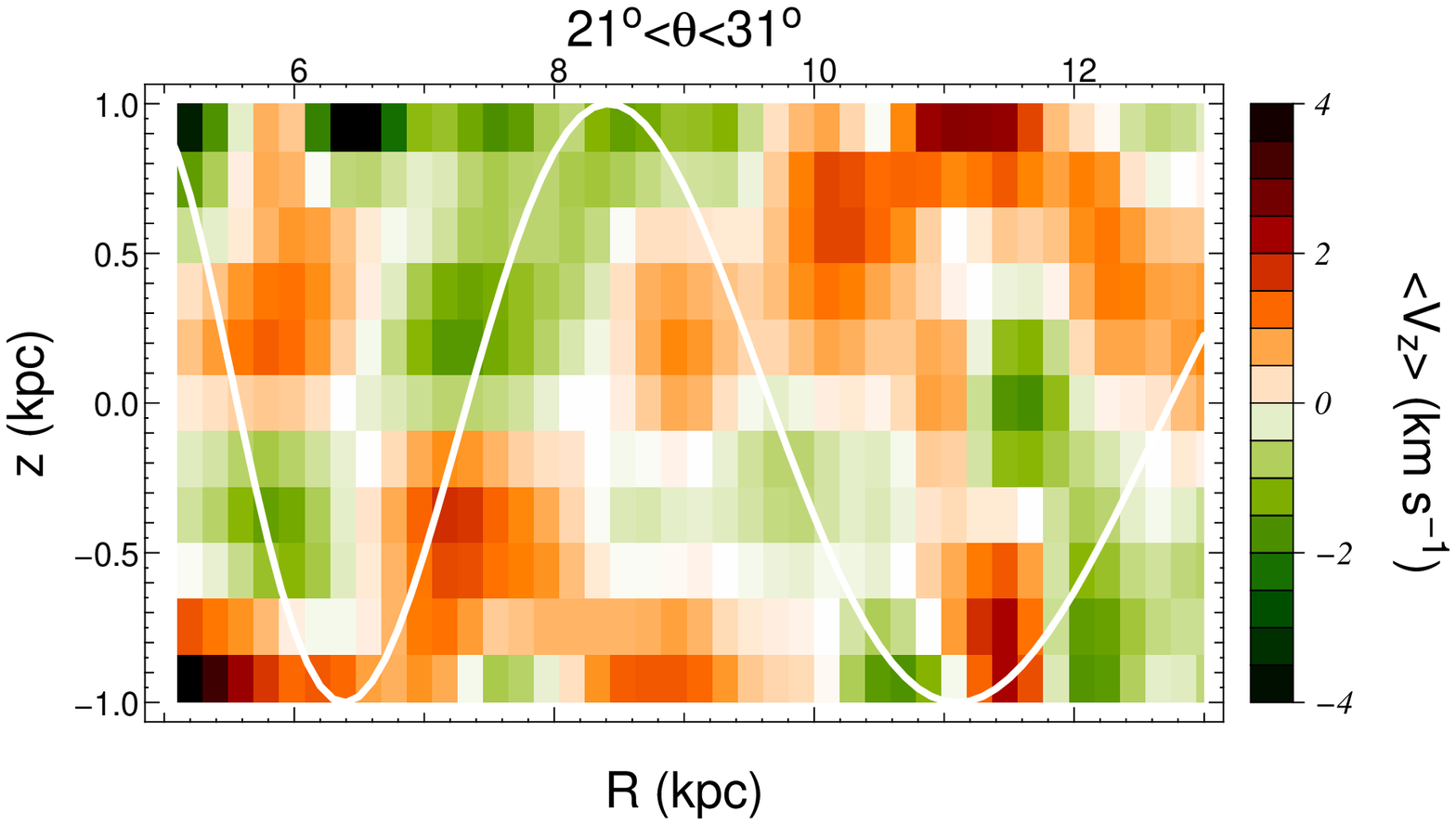}
\includegraphics[width=6cm]{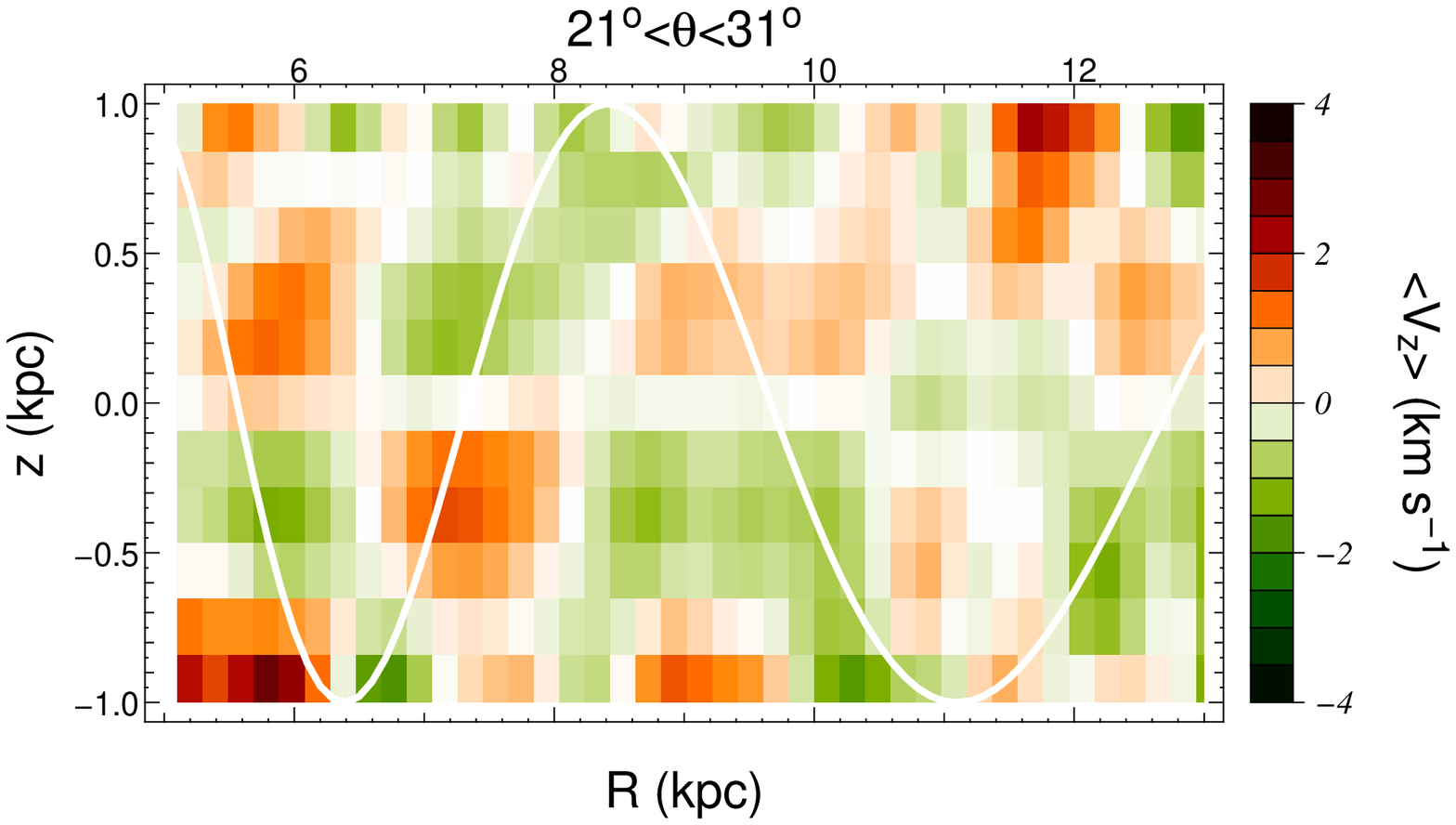}
\caption{Top-left  panel: Mean  vertical velocity  $\langle v_z  \rangle$ at
  $t=4 \,$Gyr in the meridional $(R,z)$-plane for $21^\circ<\theta<31^\circ$
  (within the frame  of the spiral). Top right: Same  at $t=5 \,$Gyr. Bottom
  left:  Same at  $t=6 \,$Gyr.  Bottom right:  Same at  the  final time-step
  $t=6.5  \,$Gyr. This  figure can  be compared  at the qualitative level to 
  Fig.~13  of  Williams et al.~(2013).}
\label{f:cartevzRZ}
\end{figure*}

\begin{figure*}
\includegraphics[width=6cm]{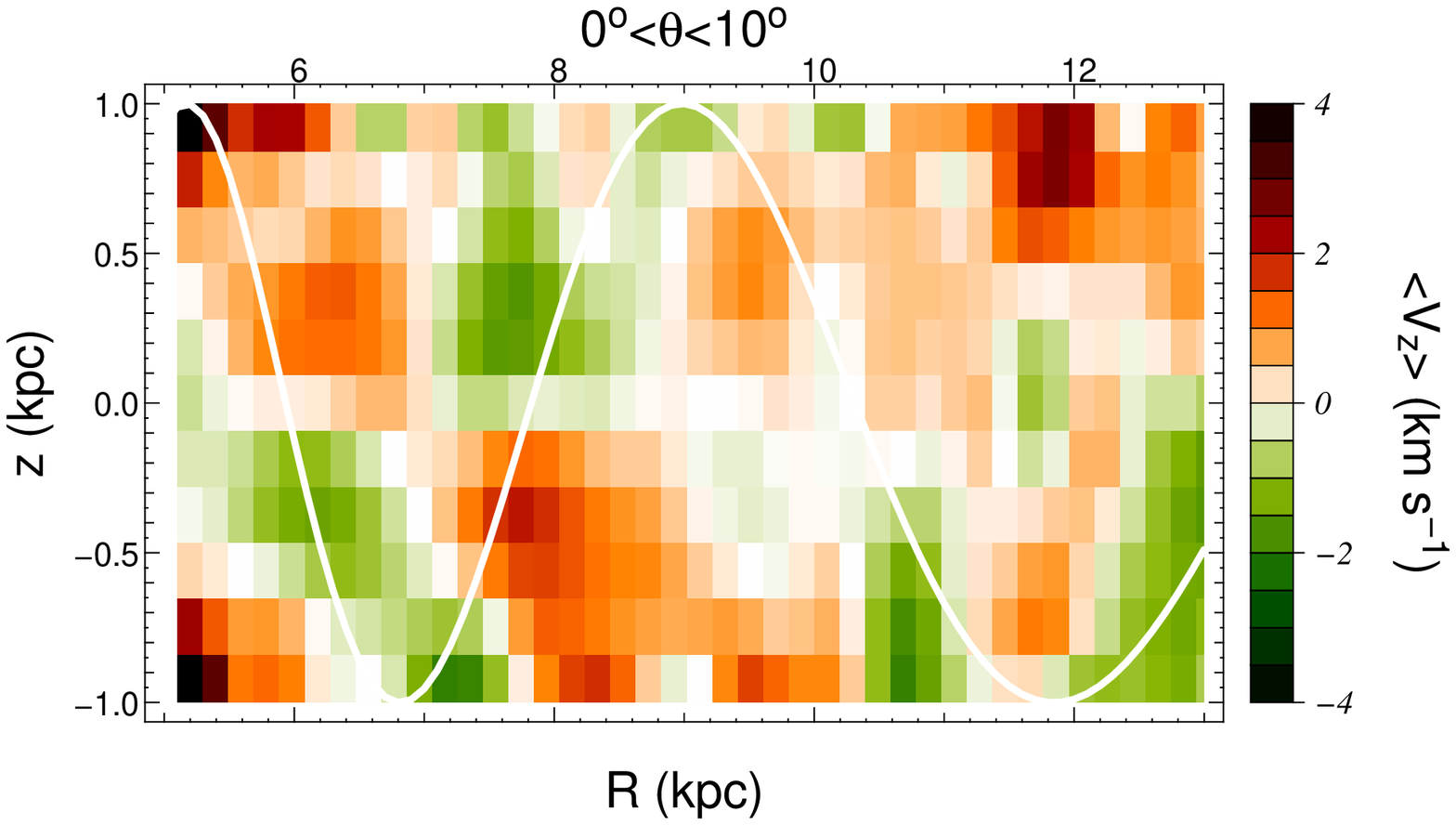}
\includegraphics[width=6cm]{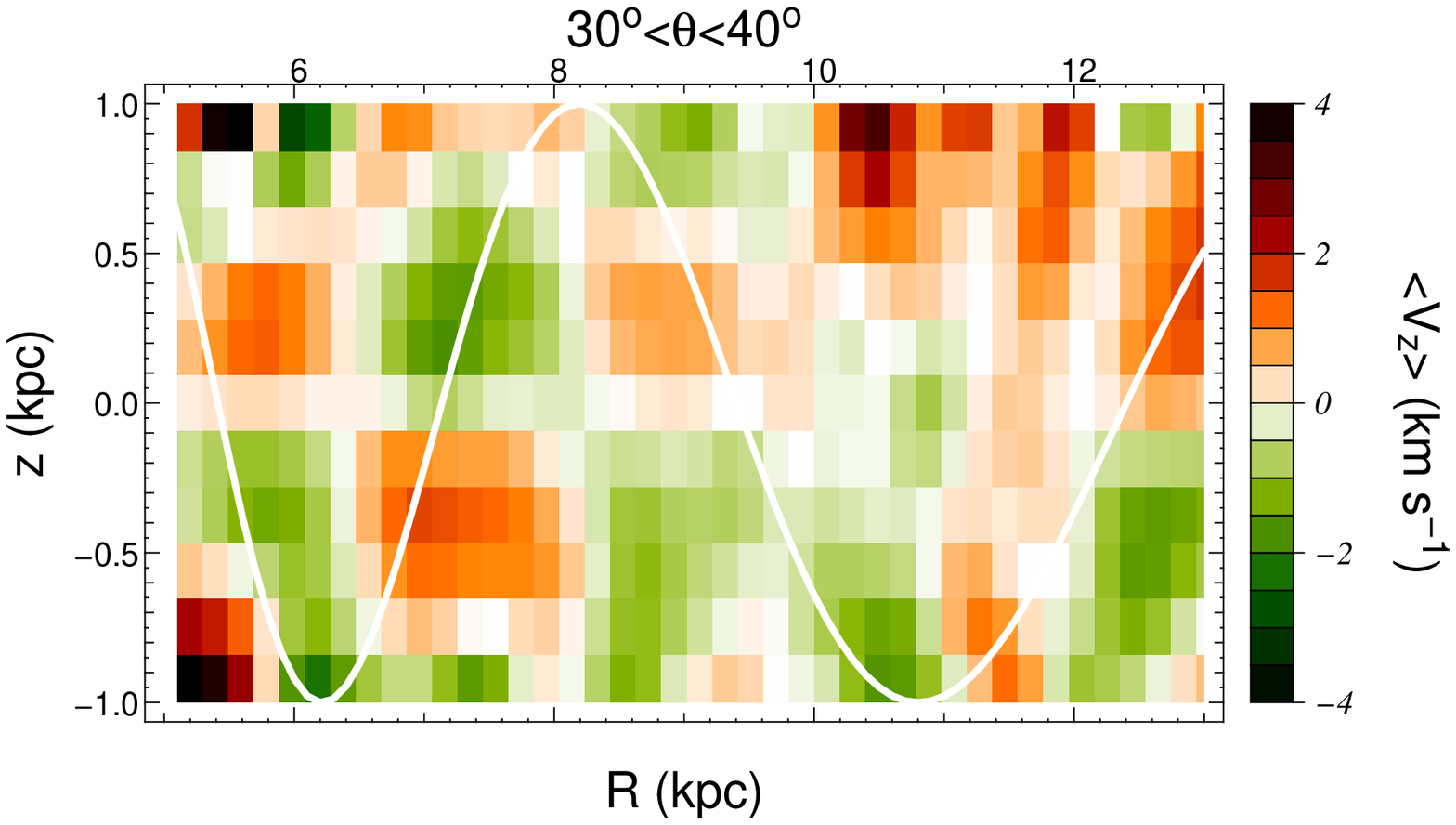}
\includegraphics[width=6cm]{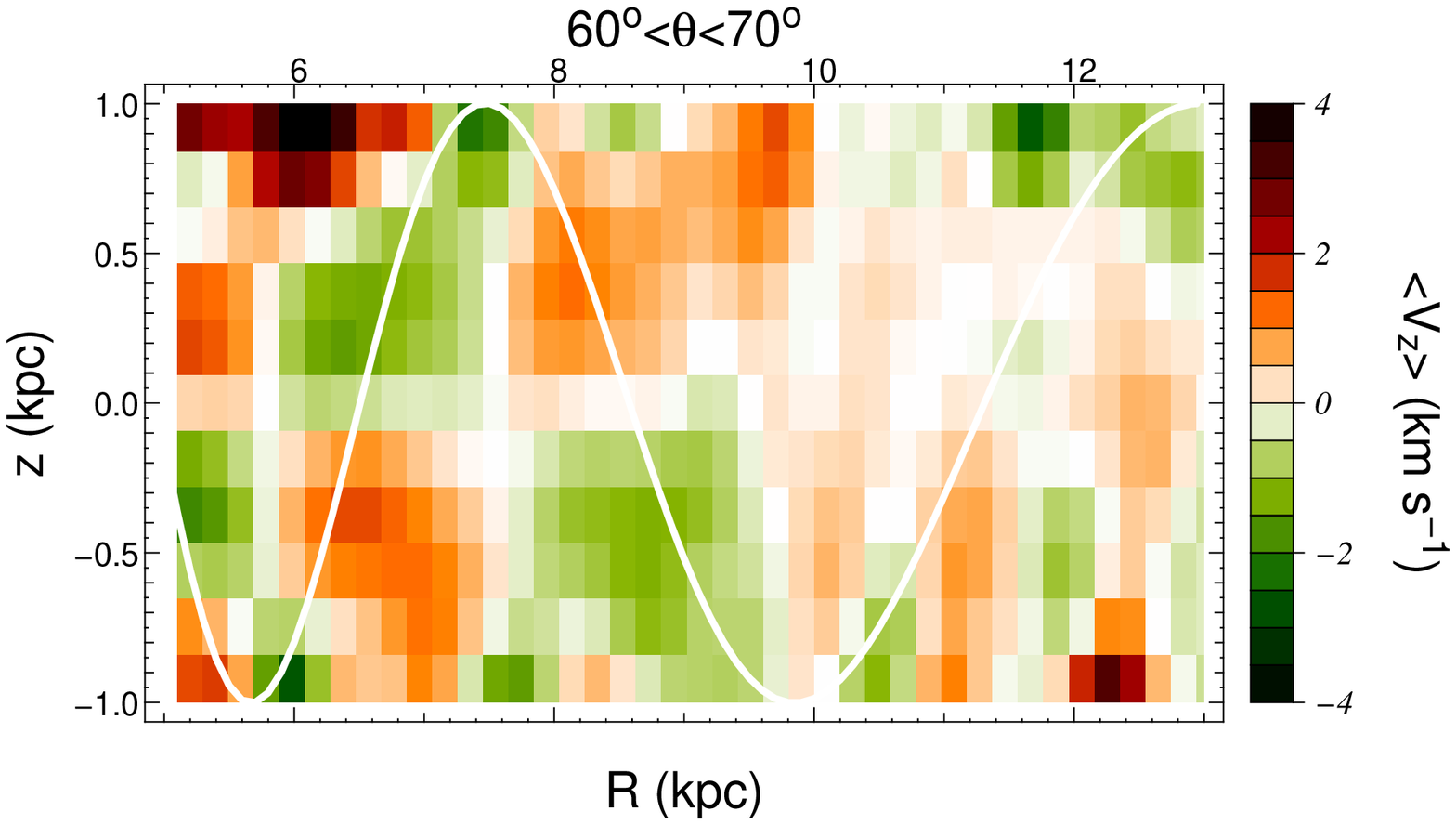}
\includegraphics[width=6cm]{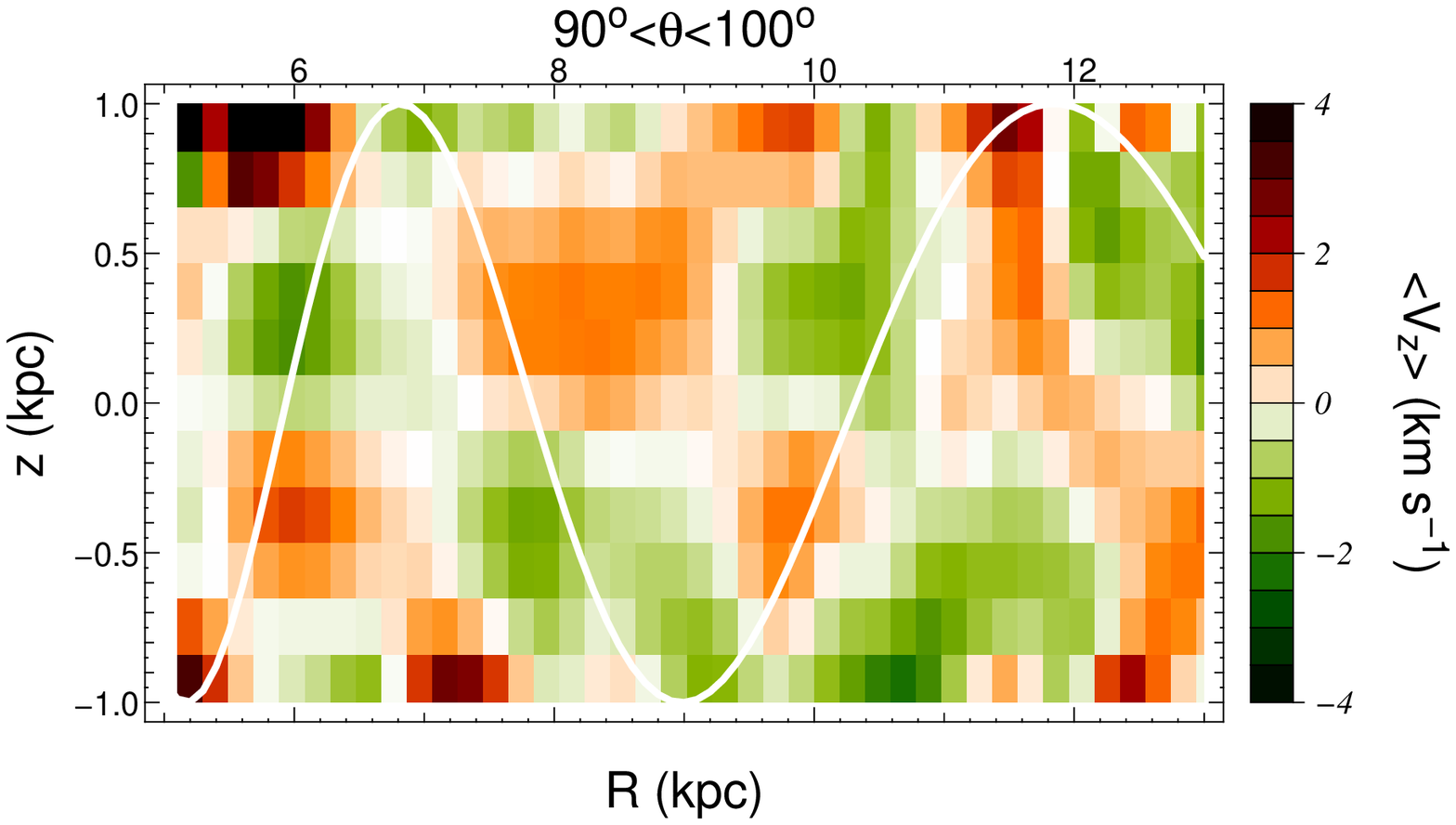}
\includegraphics[width=6cm]{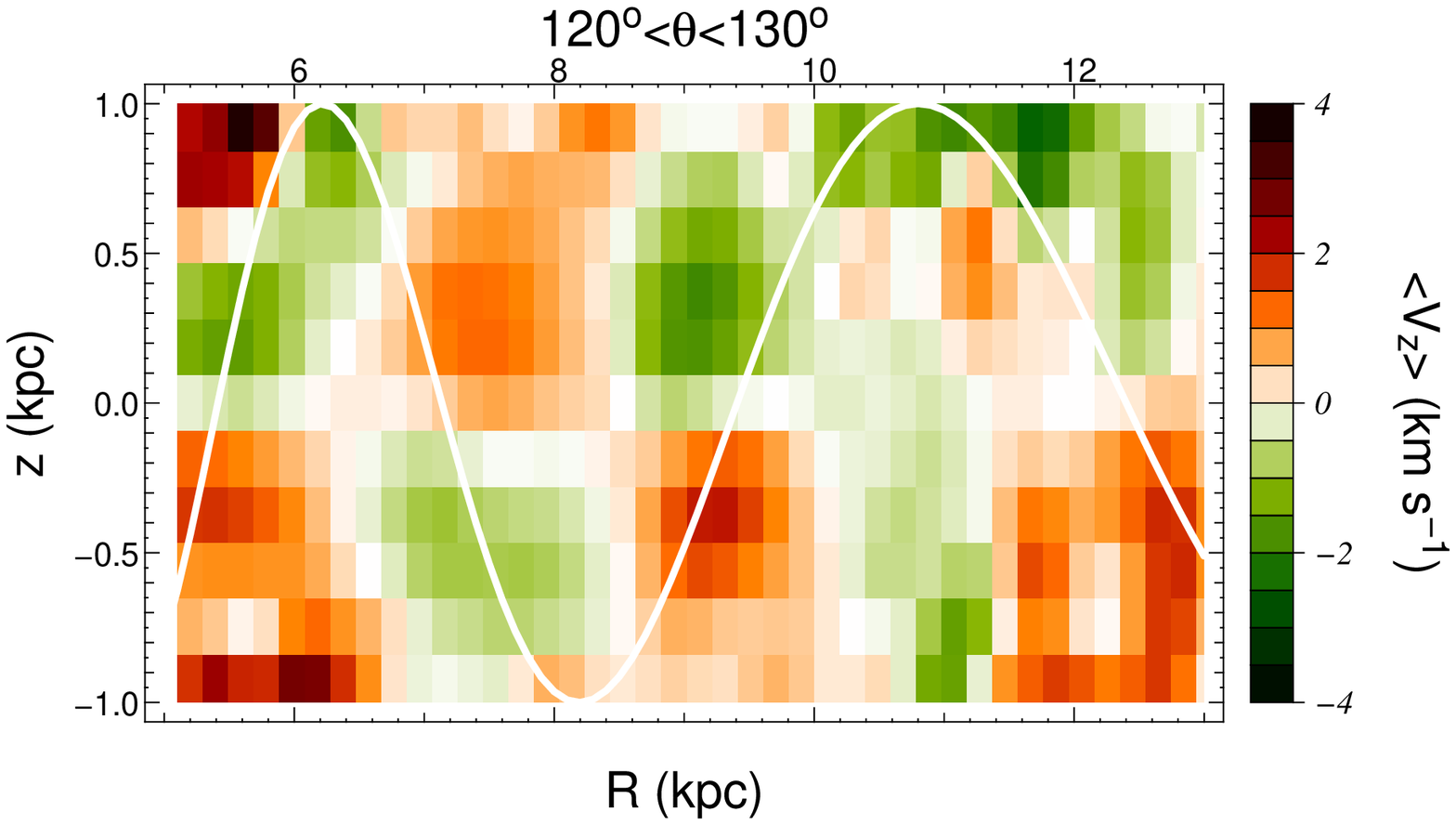}
\includegraphics[width=6cm]{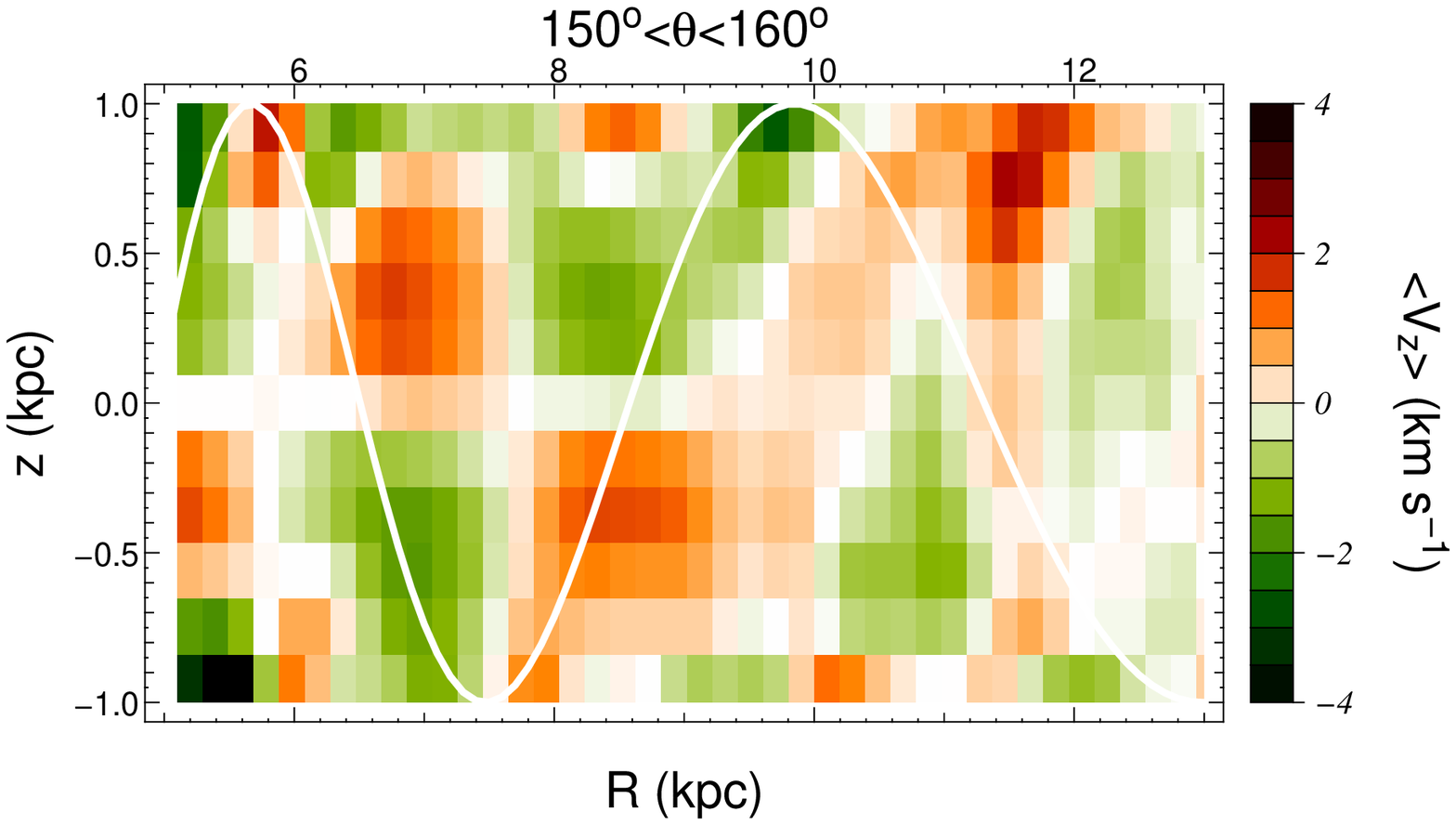}
\caption{Same as Fig.~\ref{f:cartevzRZ}, but for six different azimuths at a
  fixed time ($t=6 \,$Gyr).}
\label{f:cartevzRZ_azimuth}
\end{figure*}

\begin{figure*}
\includegraphics[width=6cm]{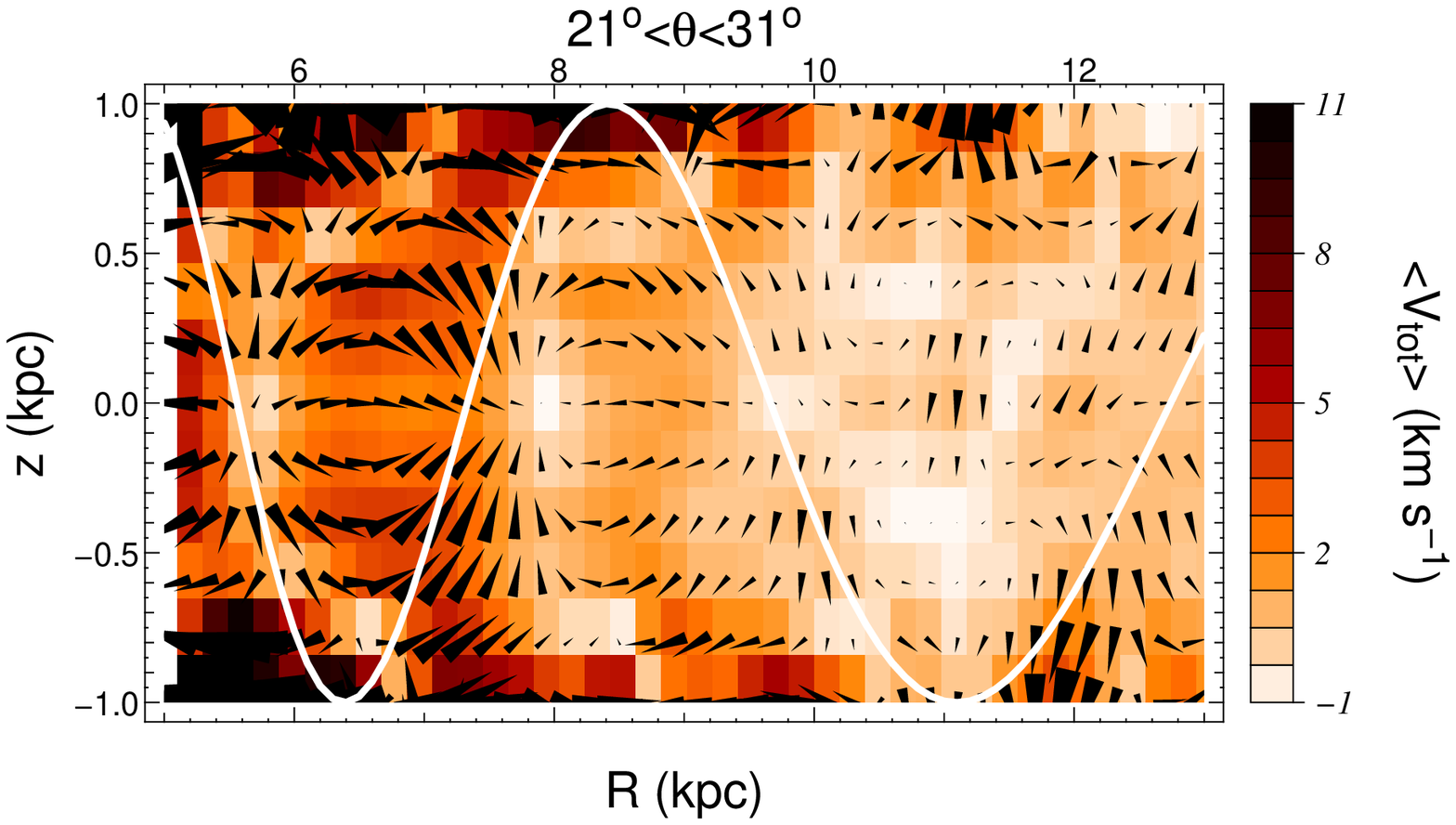}
\includegraphics[width=6cm]{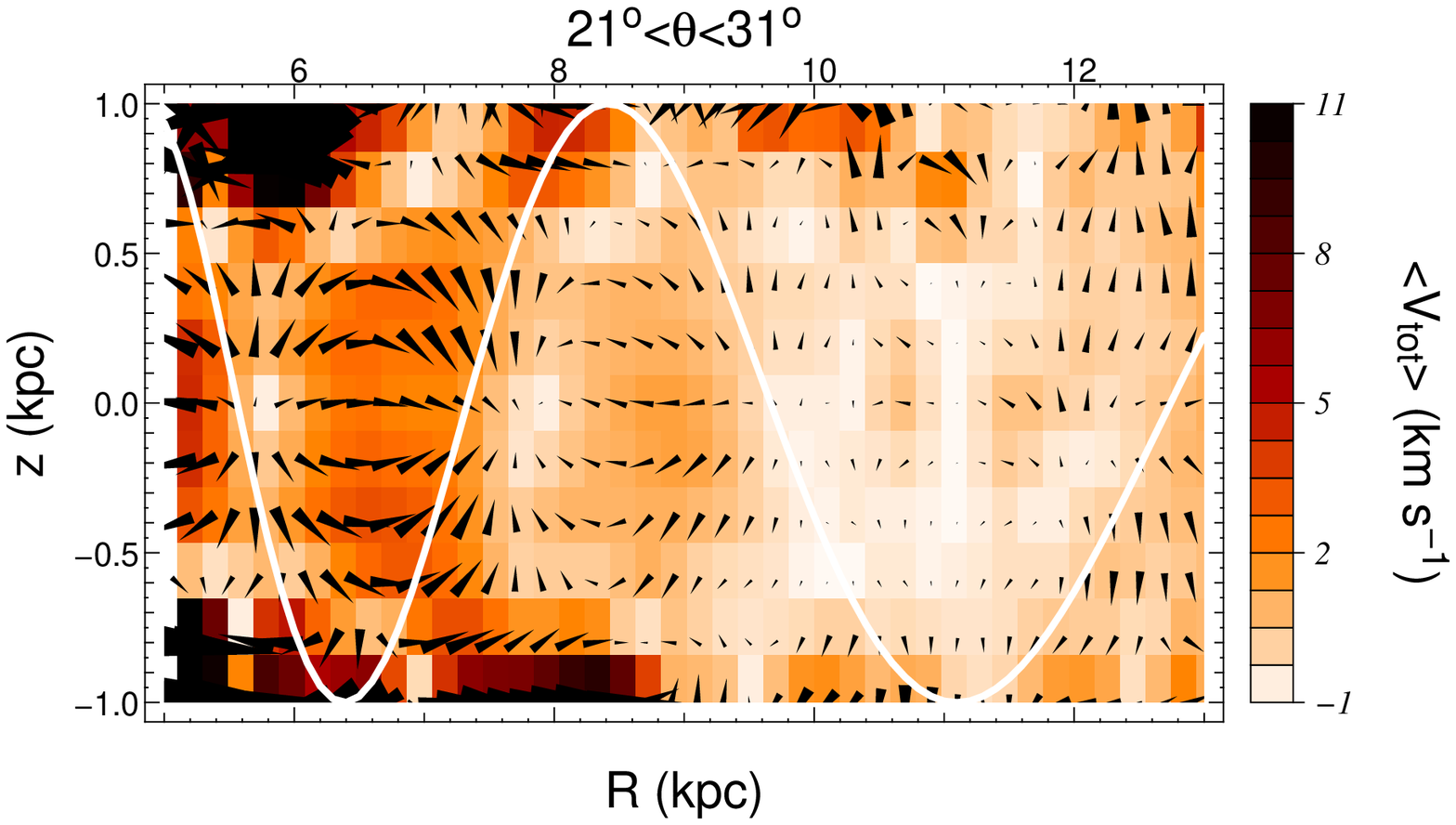}
\includegraphics[width=6cm]{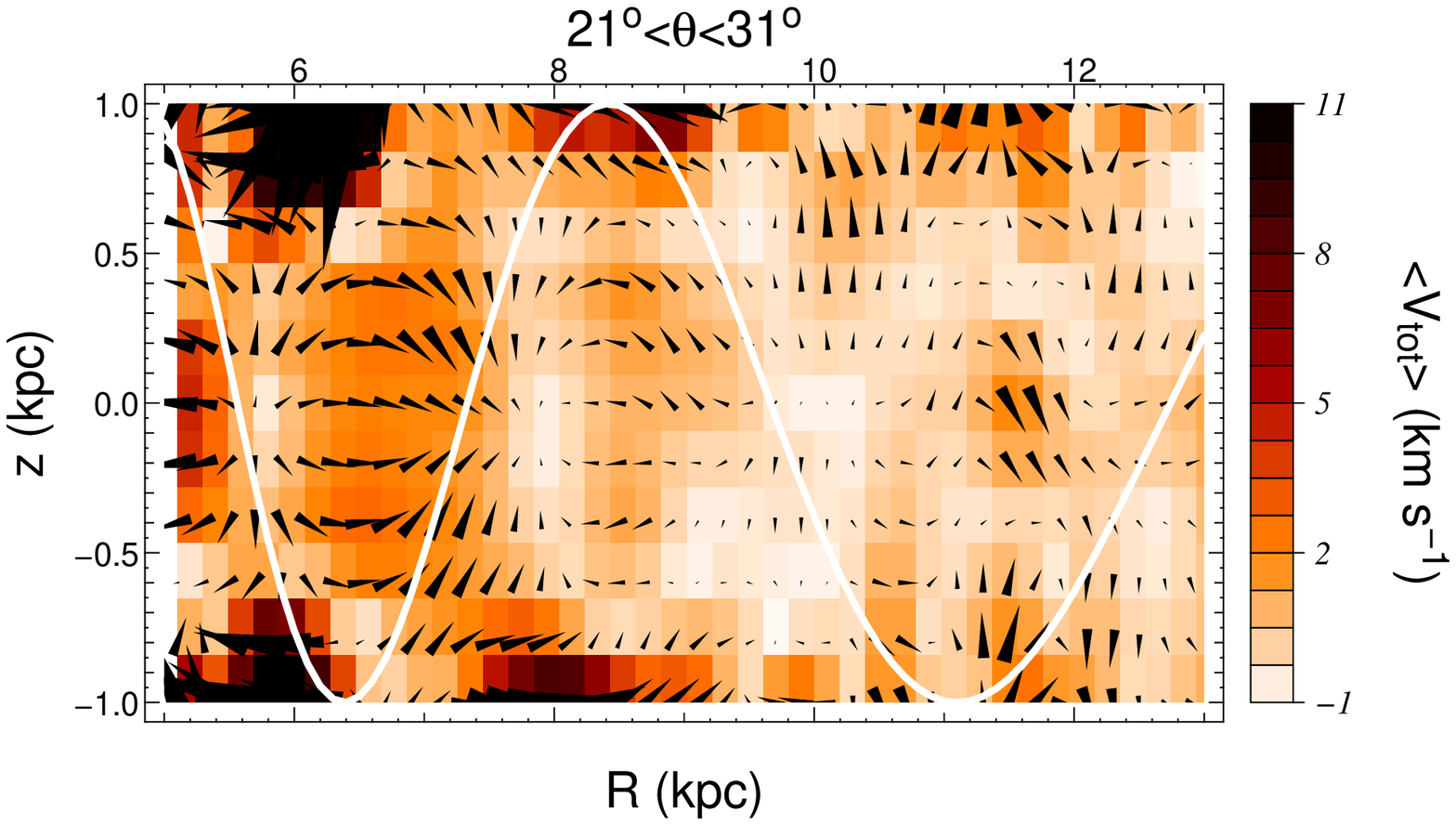}
\includegraphics[width=6cm]{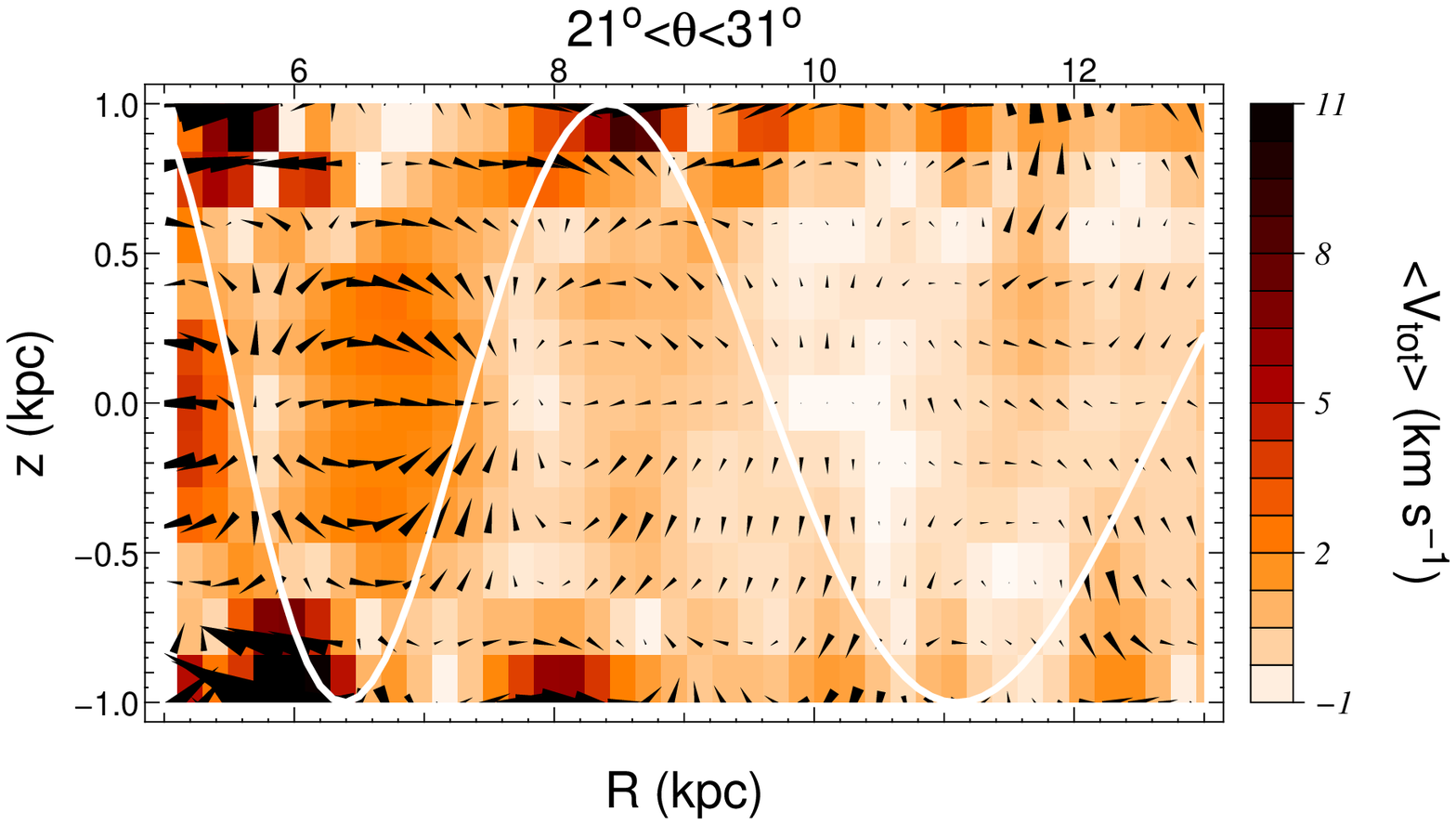}
\caption{Top-left   panel:  ``Total''   velocity  flow   in   the  meridional
  $(R,z)$-plane $\sqrt{\langle  v_R \rangle^2  + \langle v_z  \rangle^2}$ at
  $t=4 \,$Gyr for  $21^\circ<\theta<31^\circ$. Arrows indicate the direction
  of  the velocity  flow  $\vec{\langle  v \rangle}  =  \langle v_R  \rangle
  \vec{1}_R  + \langle  v_z  \rangle  \vec{1}_z$. Top  right:  Same at  $t=5
  \,$Gyr. Bottom left: Same at $t=6  \,$Gyr. Bottom right: Same at the final
  time-step $t=6.5 \,$Gyr.}
\label{f:cartevtotRZ}
\end{figure*}

\begin{figure*}
\includegraphics[width=6cm]{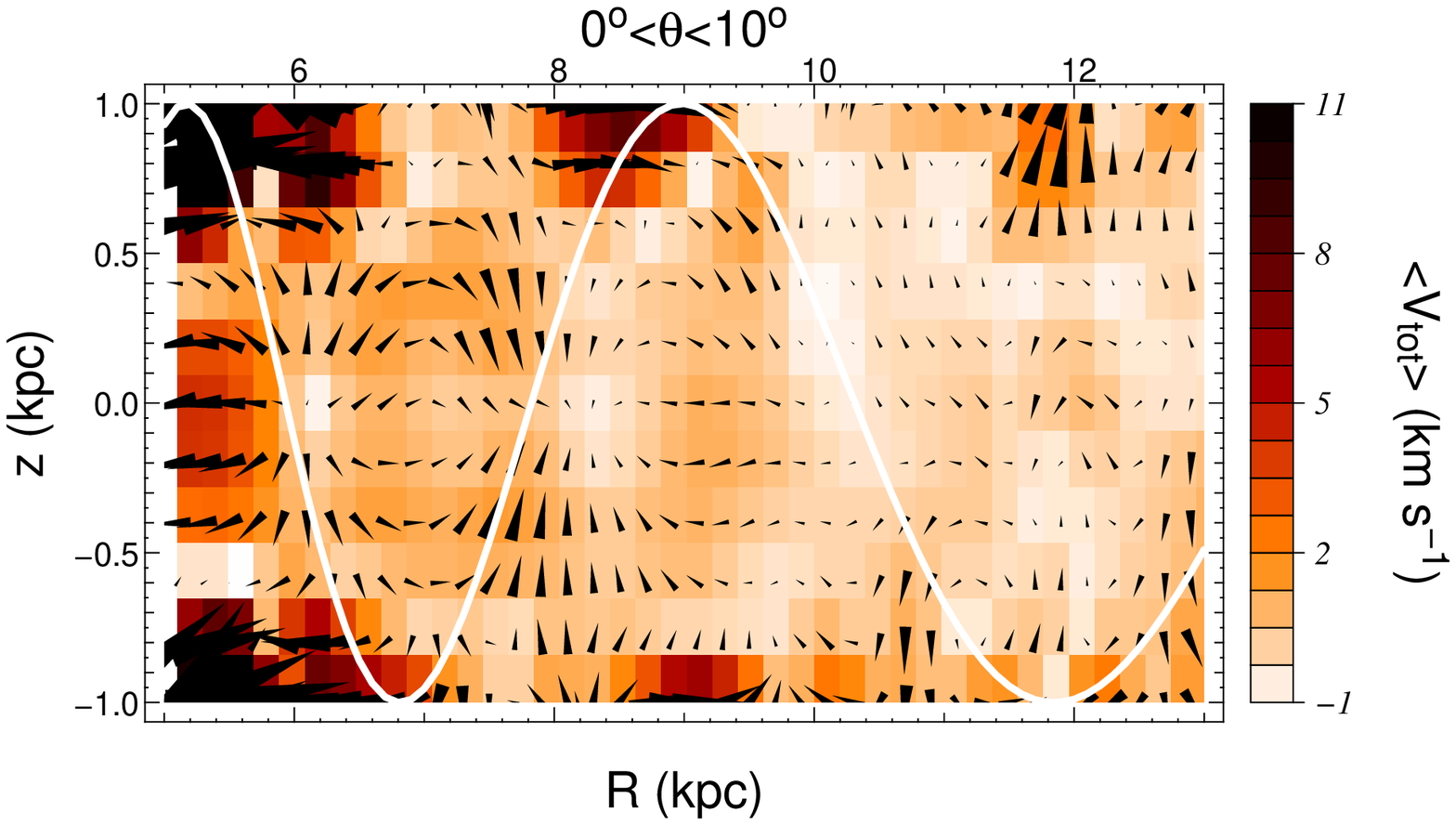}
\includegraphics[width=6cm]{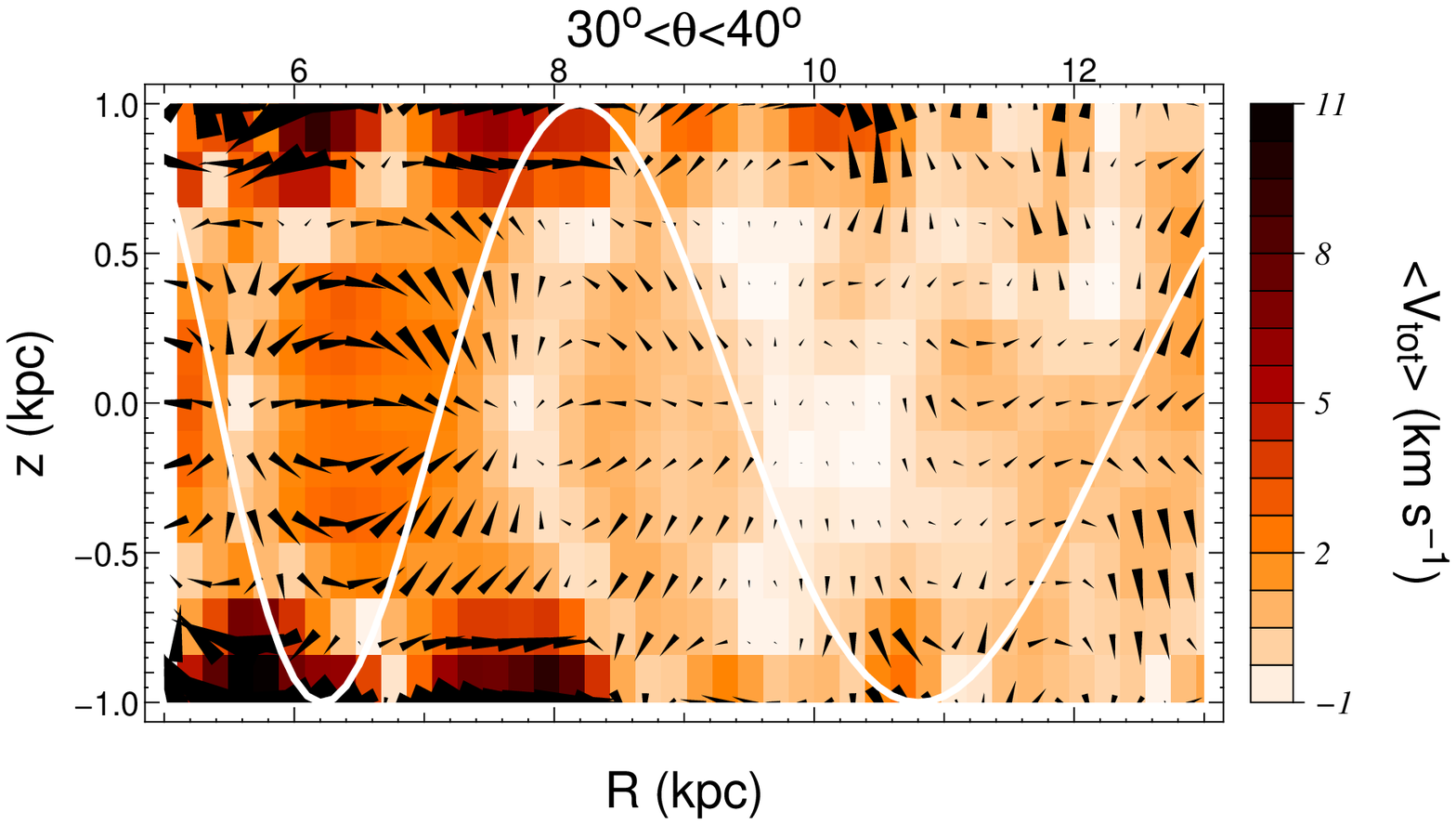}
\includegraphics[width=6cm]{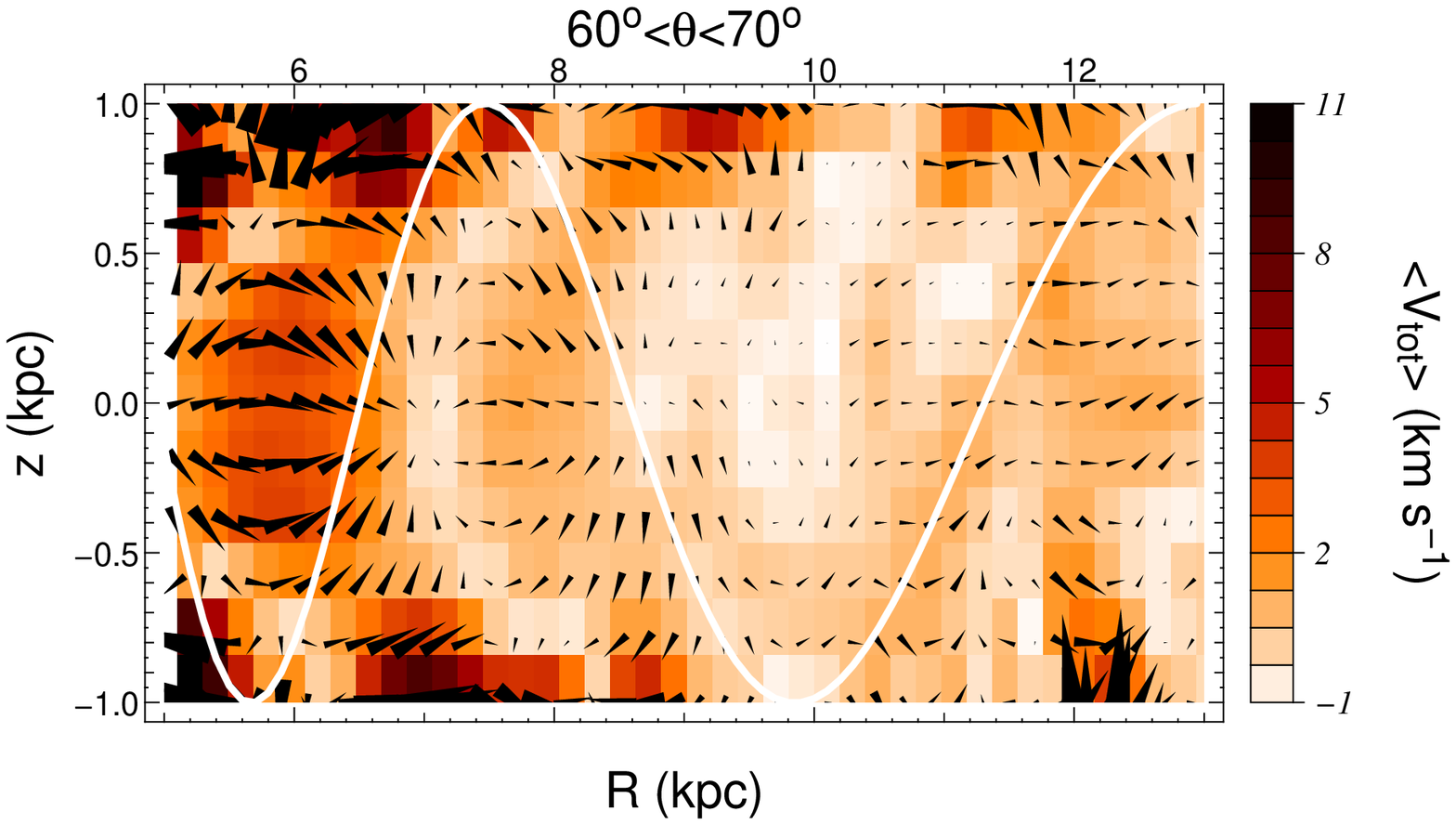}
\includegraphics[width=6cm]{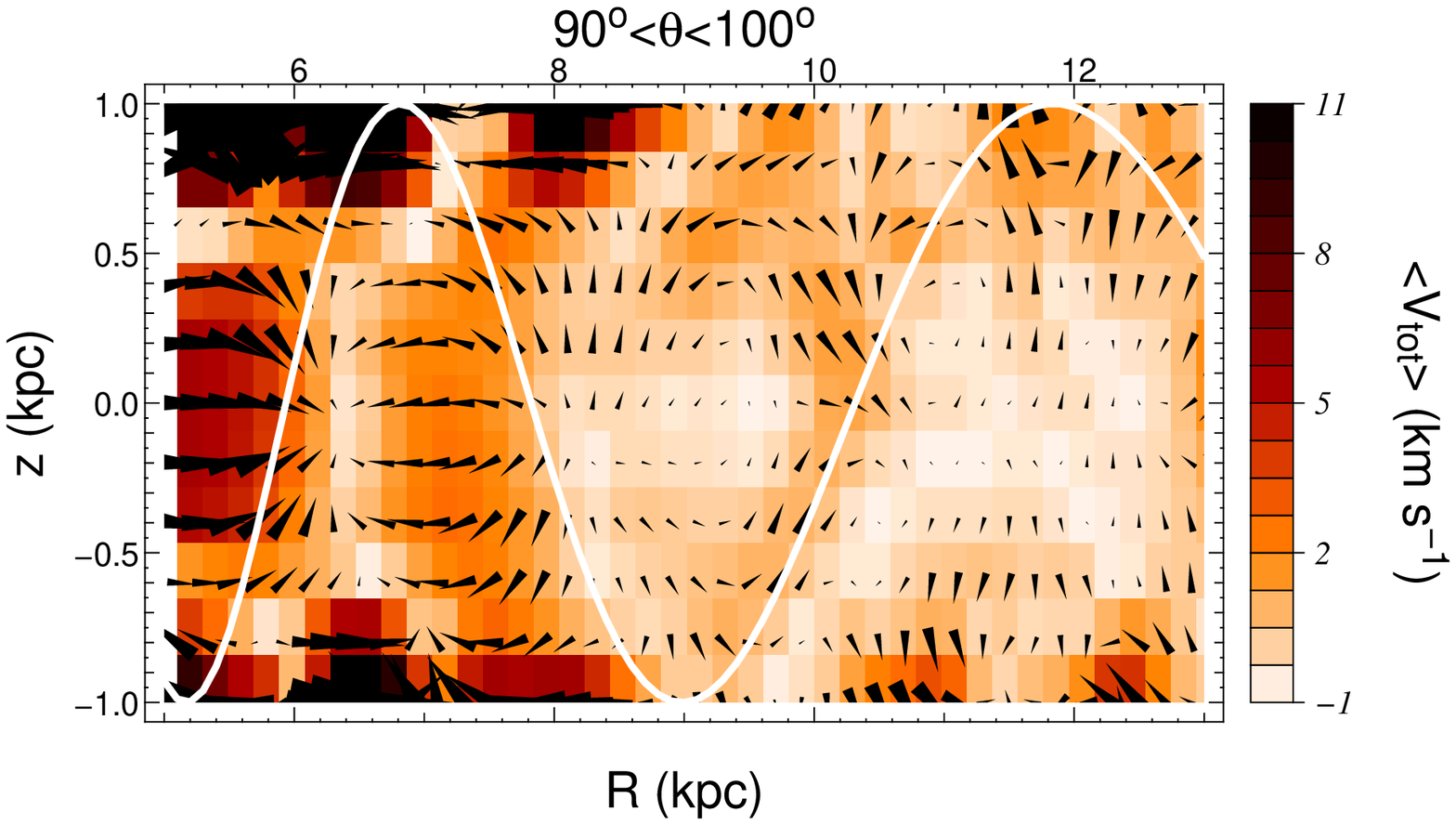}
\includegraphics[width=6cm]{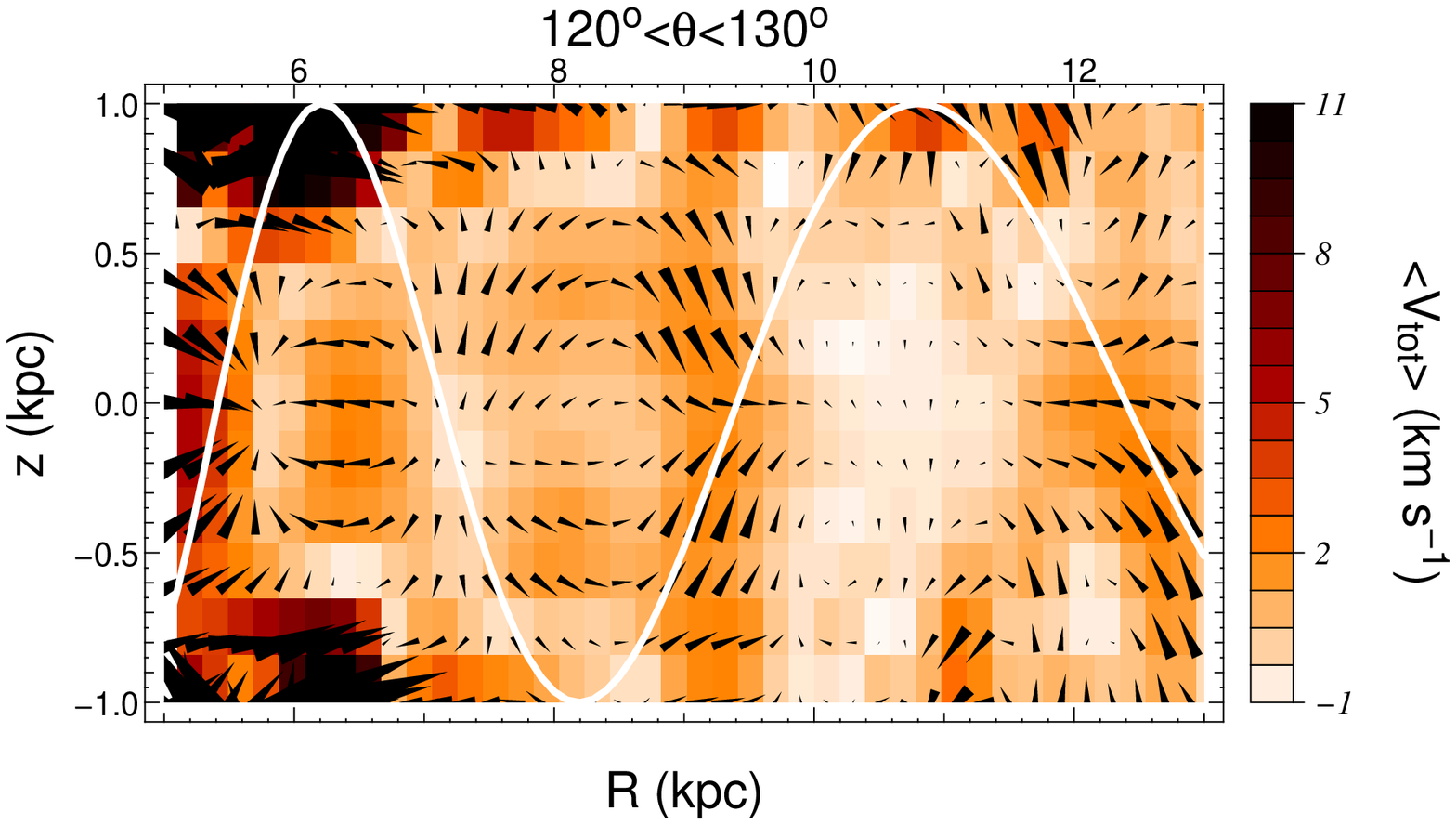}
\includegraphics[width=6cm]{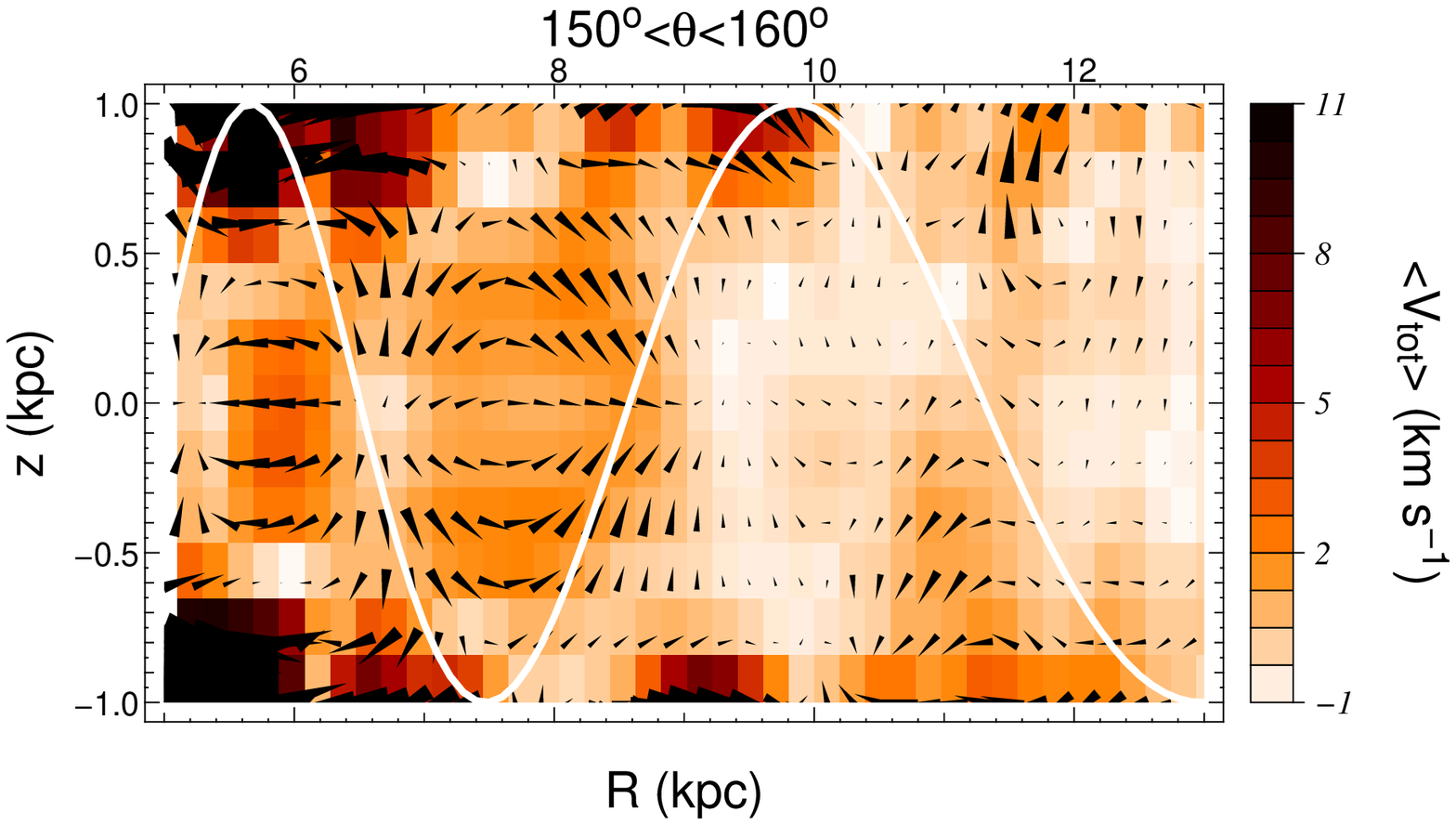}
\caption{Same as Fig.~\ref{f:cartevtotRZ}, but for six different azimuths at
  a fixed time ($t=6 \,$Gyr).}
\label{f:cartevtotRZ_azimuth}
\end{figure*}

\begin{figure*}
\includegraphics[width=8cm]{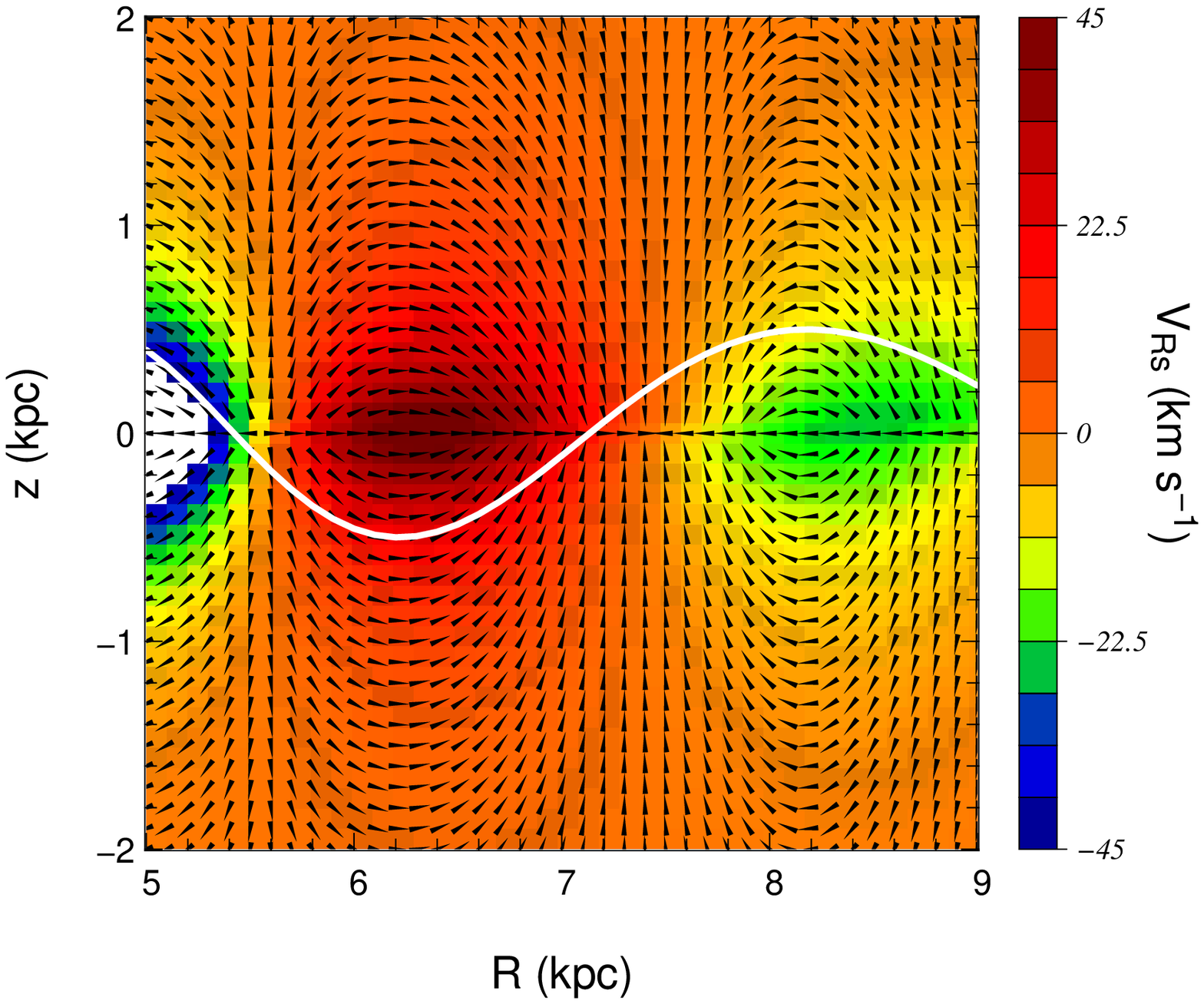}
\includegraphics[width=8cm]{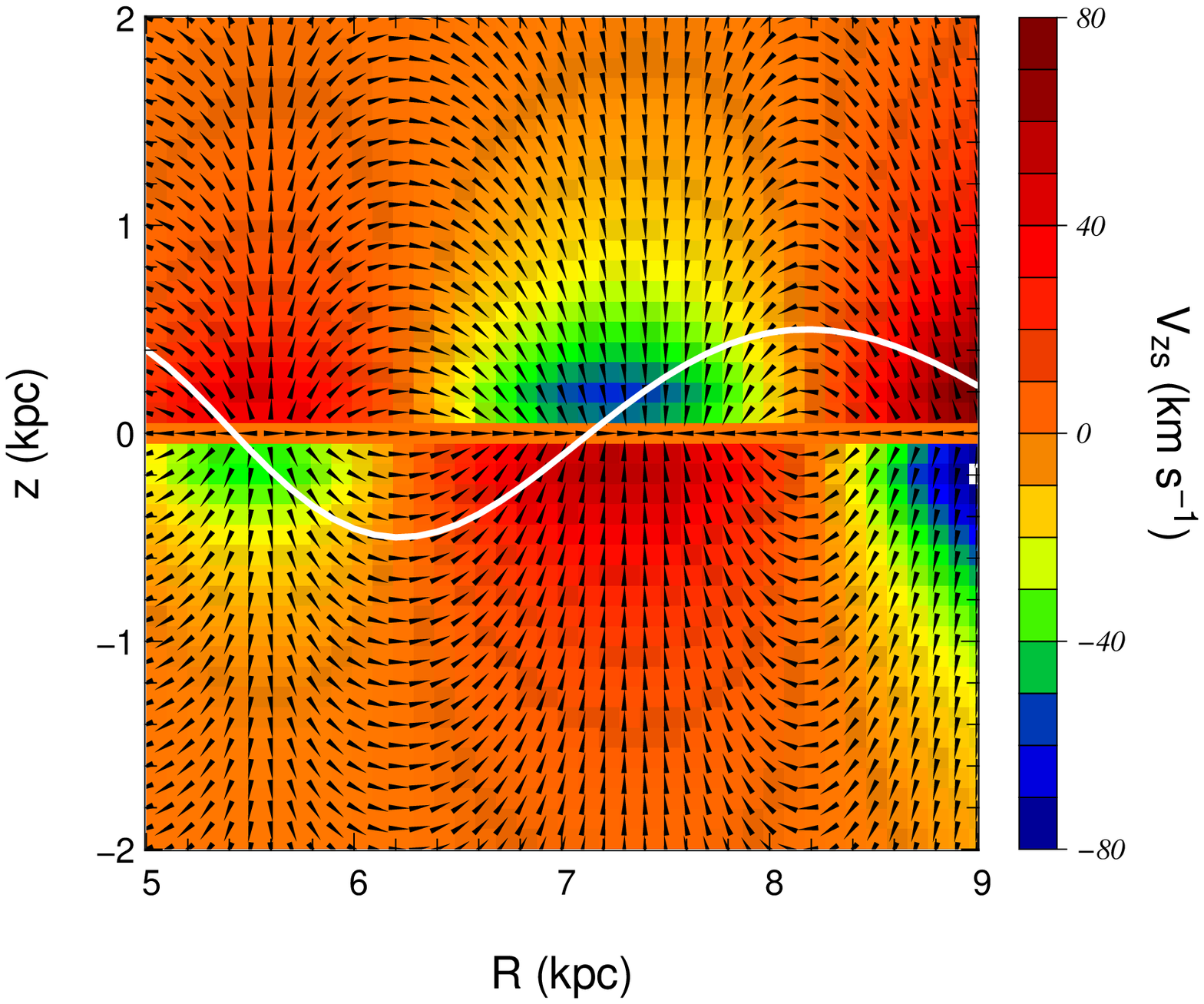}
\caption{Analytical calculation of the response  of a cold fluid to a spiral
  perturbation (Eq.~9) in terms of radial velocity $v_{Rs}$ (left panel) and
  vertical velocity  $v_{zs}$ (right panel). This velocity  flow is computed
  at  $\theta=30^\circ$ and  $t=0$  from the  solution  of linearized  Euler
  equations in Eqs.~\ref{meanv} and  \ref{va}. Arrows indicate the direction
  of  the fluid  velocity vector  in the  meridional plane.  The  white line
  indicates the  absolute value  of the spiral  potential. The shape  of the
  pattern  is the  same as  in  the simulation,  and similar  to Fig.~13  of
  Williams et al. (2013).}
\label{f:euler}
\end{figure*}

If we  now turn our  attention to  the vertical motion  of stars, we  see on
Fig.~\ref{f:sigmaz}  that the total  mean vertical  motion of  stars remains
zero  at all  times,  but that  there  is still  a  slight, but  reasonable,
vertical heating going on in the inner Galaxy.

What  is most  interesting is  to concentrate  on the  mean  vertical motion
$\langle v_z \rangle$ as a function  of position above or below the Galactic
disc.      As    can     be    seen     on     Fig.~\ref{f:cartevzRZ}    and
Fig.~\ref{f:cartevzRZ_azimuth}, while the  vertical velocities are generally
close to zero right within the plane, they are non-zero outside of it.  At a
given  azimuth within  the  frame  of the  spiral,  these non-zero  vertical
velocity patterns  are extremely stable  over time (Fig.~\ref{f:cartevzRZ}).
Within corotation the  mean vertical motion is directed  away from the plane
at the outer edge of the arm and  towards the plane at the inner edge of the
arm. The pattern of $\langle v_z \rangle$ above and below the plane are thus
mirror-images, and the  direction of the mean motion  changes roughly in the
middle of the  interarm region. This produces diagonal  features in terms of
isocontours of a given $\langle v_z \rangle$, corresponding precisely to the
observation  using RAVE  by Williams  et  al.  (2013,  see especially  their
Fig.~13), where the change of sign of $\langle v_z \rangle$ precisely occurs
in between the Perseus and Scutum main arms.

Our simulation predicts  that the $\langle v_z \rangle$  pattern is reversed
outside of corotation (beyond 12~kpc), where stars move towards the plane on
the outer edge of the arm (rather than moving away from the plane): this can
indeed  be  seen,  e.g., on  the  right  panel  of  the  second row  of  our
Fig.~\ref{f:cartevzRZ_azimuth}.   If  we  now  combine  the  information  on
$\langle  v_R \rangle$ and  $\langle v_z  \rangle$, we  can plot  the global
meridional  velocity flow  $\vec{\langle v  \rangle} =  \langle  v_R \rangle
\vec{1}_R +  \langle v_z \rangle \vec{1}_z$  on Fig.~\ref{f:cartevtotRZ} and
Fig.~\ref{f:cartevtotRZ_azimuth}.  The picture  that emerges is the following:
in the  interarm regions  located within corotation,  stars move  on average
from the inner arm to the outer  arm by going outside of the plane, and then
coming  back towards  the plane  at mid-distance  between the  two  arms, to
finally arrive back  on the inner edge of the outer  arm.  For each azimuth,
there are  thus ``source''  points, preferentially on  the outer edge  of the
arms (inside corotation,  whilst on the inner edge  outside corotation), out
of which  the mean  velocity vector flows,  while there are  ``sink'' points,
preferentially on  the inner edge  of the arms (inside  corotation), towards
which  the mean  velocity flows.   This supports  the interpretation  of the
observed   RAVE   velocity   field    of   Williams   et   al.   (2013)   as
``compression/rarefaction'' waves.

\subsection{Interpretation from linearized Euler equations}

In order to understand these  features found in the meridional velocity flow
of  our test-particle  simulation, we  now turn  to the  fluid approximation
based on linearized Euler equations,  developed, e.g., in Binney \& Tremaine
(2008, Sect.~6.2). A rigorous  analytical treatment of a quasi-static spiral
perturbation  in  a  three-dimensional  stellar  disk  should  rely  on  the
linearized Boltzmann equations, which we plan to do in full in a forthcoming
paper, but the fluid  approximation can
already give important  insights on the shape of  the velocity flow expected
in the meridional plane. In the full Boltzmann-based treatment, the velocity
flow will  be tempered by reduction  factors both in the  radial (see, e.g.,
Binney \& Tremaine 2008, Appendix K) and vertical directions.

Let us rewrite our perturber potential of Eq.~\ref{spipot} as
\begin{equation}
\Phi_s= \mathbf{Re} \lbrace \Phi_a(R,z) \X \rbrace
\end{equation}
with
\begin{equation}
\Phi_a = - A \, {\mathrm sech}^2 \left(\frac{z}{z_0} \right) \exp \Big(i \frac{m \ln(R)}{\tan p} \Big).
\end{equation}

Then  if  we write  solutions  to the  linearized  Euler  equations for  the
response of a cold fluid as
\begin{equation}
\left\{
\begin{array}{l}
v_{Rs} = \mathbf{Re} \lbrace v_{Ra}(R,z) \X
\rbrace\\
\\
v_{zs} = \mathbf{Re}  \lbrace v_{za}(R,z)) \X
\rbrace\\
\end{array}
\right.
\label{meanv}
\end{equation}
we find, following the same steps as in Binney \& Tremaine~(2008, Sect.~6.2)
\begin{equation}
\left\{
\begin{array}{l l}
v_{Ra}=& - \frac{ m (\Omega - \Omega_P)}{\Delta} k \Phi_a \\ &+ i \frac{2 \Phi_a}{\Delta} \left( \frac{2 \Omega {\rm tanh}(z/z_0)}{m (\Omega - \Omega_P) z_0} + \frac {m \Omega}{R} \right)\\
\\
v_{za} =& - \frac{ 2 i}{m (\Omega -\Omega_P) z_0} {\rm tanh}\Big( \frac{z}{z_0}\Big) \Phi_a \\
\end{array}
\right.
\label{va}
\end{equation}
where $k=m/(R \, {\rm tan \,}p)$ is the radial wavenumber and $\Delta = \kappa^2 - m^2(\Omega-\Omega_P)^2$.

If we plot  these solutions for $v_{Rs}$ and $v_{zs}$ at  a given angle (for
instance $\theta=30^\circ$)  we get  the same pattern  as in  the simulation
(Fig.~\ref{f:euler}).   Of   course,    the   velocity   flow   plotted   on
Fig.~\ref{f:euler} would in  fact be damped by a  reduction factor depending
on  both radial  and vertical  velocity dispersions  when treating  the full
linearized  Boltzmann equation,  which will  be the  topic of  a forthcoming
paper.   Nevertheless,  this   qualitative  consistency  between  analytical
results  and our  simulations is  an  indication that  the velocity  pattern
observed  by Williams  et  al.  (2013)  is  likely linked  to the  potential
perturbation by spiral arms. Interestingly, this analytical model also predicts that the radial velocity gradient should become noticeably North/South asymmetric close to corotation.

\section{Discussion and conclusions}

In recent years,  various large spectroscopic surveys have  shown that stars
of  the Milky  Way  disc exhibit  non-zero  mean velocities  outside of  the
Galactic  plane in  both the  Galactocentric radial  component  and vertical
component of the  mean velocity field (e.g., Siebert  et al.  2011b; Williams
et al.  2013; Carlin  et al. 2013). While it is clear  that such a behaviour
could be due to a large combination of factors, we investigated here whether
spiral arms are able to play a role in these  observed patterns. For this
purpose, we investigated the orbital  response of a test population of stars
representative of the old thin disc to a stable spiral perturbation. This is
done   using  a   test-particle  simulation   with  a   background  potential
representative of the Milky Way.

We found non-zero velocities both  in the Galactocentric radial and vertical
velocity components.  Within  the rotating frame of the  spiral pattern, the
location of  these non-zero  mean velocities in  both components  are stable
over time, meaning  that the response to the  spiral perturbation is stable.
Within corotation,  the mean  $\langle v_R \rangle$  is negative  within the
arms (mean radial  motion towards the Galactic centre)  and positive (radial
motion  towards the anticentre)  between the  arms. Outside  corotation, the
pattern is reversed, as expected  from the Lin-Shu density
wave theory  (Lin \& Shu  1964). On the  other hand, even though  the spiral
perturbation  of the potential  is very  thin, the  radial velocity  flow is
still  strongly affected above  the Galactic  plane.  Up  to five  times the
scale-height of  the spiral  potential, there are  no strong  asymmetries in
terms of radial  velocity, but above these heights, the  trend in the radial
velocity flow is reversed. This  means that asymmetries could be observed in
surveys covering different volumes above and below the Galactic plane. Also,
forthcoming surveys like Gaia, 4MOST, WEAVE  will be able to map this region
of the disc  of the Milky Way  and measure the height at  which the reversal
occurs.  Provided   this  measurement  is  successful,  it   would  give  a
measurement of the scale height of the spiral potential.

In terms of vertical velocities, within corotation, the mean vertical motion
is directed away  from the plane at  the outer edge of the  arms and towards
the  plane at  the  inner edge  of the  arms.  The patterns  of $\langle  v_z
\rangle$ above and  below the plane are thus  mirror-images (see e.g. Carlin
et al.  2013). The direction of  the mean vertical motion changes  roughly in the
middle of  the interam region. This  produces diagonal features  in terms of
isocontours of  a given  $\langle v_z \rangle$,  as observed by  Williams et
al. (2013). The picture that emerges from our simulation is one of ``source''
points of the  velocity flow in the meridional  plane, preferentially on the
outer edge of the arms (inside  corotation, whilst on the inner edge outside
corotation), and of ``sink'' points,  preferentially on the inner edge of the
arms (inside corotation), towards which the mean velocity flows.

We have then shown that this qualitative structure of the mean velocity field 
is also the behaviour of the analytic solution to linearized Euler equations for a toy model of a cold fluid in response to a spiral  perturbation.  In a more realistic analytic model, this fluid velocity  would in fact be  damped by a reduction  factor depending on both  radial  and  vertical  velocity  dispersions when  treating  the  full linearized  Boltzmann equation.

In a next step,  the features
found in the  present test-particle simulations will also  be checked for in
fully  self-consistent simulations with  transient spiral  arms, to check
whether non-zero mean vertical motions as found here are indeed generic. The
response  of the gravitational  potential itself  to these  non-zero motions
should also  have an  influence on the  long-term evolution of  the velocity
patterns found here, in the form of e.g. bending and corrugation waves. The effects of multiple spiral patterns (e.g., Quillen et al. 2011) and of the bar (e.g., Monari et al. 2013, 2014) should also have an influence on the global velocity fiel and on its amplitude. Once all these different dynamical effects and their combination will be fully understood, a full quantitative comparison with present and future datasets in 3D will be the next step.

The present work on the orbital response  of the thin disc to a small spiral
perturbation by  no means implies that  no external perturbation  of the Milky
Way disc happened in the recent  past, by e.g. the Sagittarius dwarf (e.g., Gomez et al. 2013). Such a perturbation  could  of course  be  responsible  for  parts of  the  velocity structures   observed  in  various   recent  large   spectrosocpic  surveys.
For instance, concerning the  important north-south asymmetry spotted in stellar densities at  relatively large heights  above the  disc, spiral
arms are less likely to play  an important role. Nevertheless, any external perturbation will also excite a spiral wave, so that understanding the dynamics of spirals is also fundamental to understanding the effects of an external perturber. The qualitative similarity between  our  simulation  (e.g.,  Fig.~\ref{f:cartevzRZ}), as  well  as  our analytical estimates  for the fluid  approximation (Fig.~\ref{f:euler}), and the velocity pattern observed  by Williams  et al.   (2013, their Fig.~13)  indicates that
spiral arms  are likely to play a non-negligible role in the observed  velocity pattern of our ``wobbly Galaxy''.

\bibliographystyle{mn2e}

\begin{thebibliography}{}  
\bibitem{Antoja} Antoja T., Valenzuela O., Pichardo B., et al., 2009, ApJ, 700, L78
\bibitem{Antoja2} Antoja T., Figueras F., Romero-Gómez M., et al., 2011, MNRAS, 418, 1423
\bibitem{Barros} Barros D., L\'epine J., Junqueira T., 2013, MNRAS, 435, 2299
\bibitem{Bienayme} Bienaym\'e O., S\'echaud N., 1997, A\&A, 323, 781
\bibitem{BT} Binney J., Tremaine S., 2008, Galactic Dynamics, Princeton University Press
\bibitem{Binney} Binney J., 2013, New Astronomy Reviews, 57, 29
\bibitem{Binrave} Binney J., Burnett B., Kordopatis G., et al., 2013, arXiv:1309.4285
\bibitem{Bovy1} Bovy J., Allende Prieto C., Beers T., et al., 2012, ApJ, 759, 131
\bibitem{Bovy2} Bovy J., Rix H.-W., 2013, arXiv:1309.0809
\bibitem{Carlin} Carlin J.L., DeLaunay J., Newberg H.~J., et al., 2013, ApJ, 777, L5
\bibitem{chereul1998} Chereul E., Cr\'ez\'e M., Bienaym\'e O., 1998, A\&A,
  340,384
\bibitem{chereul1999} Chereul E., Cr\'ez\'e M., Bienaym\'e O., 1999, A\&AS,
  135, 5
\bibitem{dehnen1998} Dehnen W., 1998, AJ, 115, 2384 
\bibitem{galpot} Dehnen W., Binney J., 1998, MNRAS, 294, 429 
\bibitem{dehnen2000} Dehnen W., 2000, AJ, 119, 800 
\bibitem{desim} De Simone R., Wu X., Tremaine S., 2004, MNRAS, 350, 627
\bibitem{fam05} Famaey B., Jorissen A., Luri X., et al., 2005, A\&A, 430, 165 
\bibitem{fam07} Famaey B., Pont F., Luri X., et al., 2007, A\&A 461, 957
\bibitem{fam08} Famaey B., Siebert A., Jorissen A., 2008, A\&A, 483, 453 
\bibitem{Feldmann} Feldmann R., Spolyar D., 2013, arXiv:1310.2243
\bibitem{Gomez} Gomez F., Minchev I., O'Shea B., et al., 2013, MNRAS, 429, 159
\bibitem{Kaasa1} Kaasalainen M., Binney J., 1994, MNRAS, 268, 1033
\bibitem{Kaasa2} Kaasalainen M., 1994, MNRAS, 268, 1041
\bibitem{Kaasa3} Kaasalainen M., 1995, Phys. Rev. E, 52, 1193
\bibitem{Kordopatis2013} Kordopatis G., Gilmore G., Steinmetz M., et al.,
  2013, 146, 134
\bibitem{Kuijken} Kuijken K., Tremaine S., 1994, ApJ, 421, 178
\bibitem{Lepine} L\'epine J., Cruz P., Scarano S., et al., 2011, MNRAS, 417, 698
\bibitem{Linshu} Lin C.C., Shu F.H., 1964, ApJ, 140, 646
\bibitem{McMillan2010} McMillan P.J., Binney J., 2010, MNRAS, 402, 934
\bibitem{McMillan} McMillan P.J., 2013, MNRAS, 430, 3276
\bibitem{Min2007} Minchev I., Nordhaus J., Quillen A., 2007, ApJ, 664, L31
\bibitem{Min2010} Minchev I., Famaey B., 2010, ApJ, 722, 112
\bibitem{Minboily} Minchev I., Boily C., Siebert A., Bienaym\'e O., 2010, MNRAS, 407, 2122 
\bibitem{Min2012} Minchev I., Famaey B., Quillen A., et al., 2012, A\&A, 548, A126
\bibitem{Monari} Monari G., Antoja T., Helmi A., 2013, arXiv:1306.2632
\bibitem{Monari2} Monari G., Helmi A., Antoja T., Steinmetz M., 2014, arXiv:1402.4479
\bibitem{Olling} Olling R., Dehnen W., 2003, ApJ, 599, 275
\bibitem{Patsis} Patsis P.A., Grosb{\o}l P., 1996, A\&A, 315, 371
\bibitem{Pompeia} Pomp\'eia L., Masseron T., Famaey B., et al., 2011, MNRAS, 415, 1138
\bibitem{quillen2005} Quillen A., Minchev I., 2005, AJ, 130, 576
\bibitem{Quillen} Quillen A., Dougherty J., Bagley M., et al., 2011, MNRAS, 417, 762
\bibitem{Reid} Reid M., Menten K., Zheng X., et al. 2009, ApJ, 700, 137
\bibitem{Roskar} Roskar R., Debattista V., Quinn T., Wadsley J., 2012, MNRAS, 426, 2089
\bibitem{Schonrich} Sch\"onrich R., 2012, MNRAS, 427, 274
\bibitem{Sellwood} Sellwood J., 2013, Rev. Mod. Phys., arXiv:1310.0403
\bibitem{Shu} Shu F.H., 1969, ApJ, 158, 505
\bibitem{Siebert2011a} Siebert A., Williams M.E.K., Siviero A., et al.,
  2011a, AJ, 141, 187 
\bibitem{Siebert1} Siebert A., Famaey B., Minchev I., et al., 2011b, MNRAS, 412, 2026
\bibitem{Siebert2} Siebert A., Famaey B., Binney J., et al., 2012, MNRAS, 425, 2335
\bibitem{Smith} Smith M., Whiteoak S.~H., Evans N.~W., 2012, ApJ, 746, 181
\bibitem{Steinmetz2006} Steinmetz M., Zwitter T., Siebert A., et al., 2006,
  AJ, 132, 1645
\bibitem{Widrow} Widrow L., Gardner S., Yanny B., et al., 2012, ApJ, 750, L41
\bibitem{Williams} Williams M., Steinmetz M., Binney J., et al., 2013, MNRAS, 436, 101
\bibitem{Yanny} Yanny B., Gardner S., 2013, ApJ, 777, 91
\bibitem{Zwitter2008} Zwitter T., Siebert A., Munari U., et al., 2008, AJ,
  136, 421
\end{thebibliography}

%\bsp  
  
\label{lastpage}  

\end{document}